\documentclass{article}

\usepackage{graphicx,epsfig,color}
\usepackage{cite}
\usepackage[english]{babel}
\usepackage{amssymb,amsfonts,amsmath}
\usepackage{verbatim}
\newcommand{\cA}{{\cal A}}  \newcommand{\cB}{{\cal B}}

  \newcommand{\cL}{{\cal L}}
  \newcommand{\cN}{{\cal N}}
\newcommand{\cO}{{\cal O}}

  \newcommand{\cV}{{\cal V}}

\newcommand{\be}{\begin{equation}} \newcommand{\ee}{\end{equation}}
\newcommand{\bea}{\begin{eqnarray}} \newcommand{\eea}{\end{eqnarray}}
\newcommand{\beann}{\begin{eqnarray*}}  \newcommand{\eeann}{\end{eqnarray*}}
\newcommand{\bfig}{\begin{figure}} \newcommand{\efig}{\end{figure}}
\newcommand{\ba}{\begin{array}} \newcommand{\ea}{\end{array}}
\newcommand{\bcen}{\begin{center}} \newcommand{\ecen}{\end{center}}
\newcommand{\btab}{\begin{tabular}} \newcommand{\etab}{\end{tabular}}

\def\tr{\operatorname{tr\:}}

\newcommand{\vev}[1]{\left\langle{#1}\right\rangle}

\newtheorem{Proposition}{Proposition}[section]

\newtheorem{Theorem}{Theorem}[section]
\newtheorem{Lemma}{Lemma}[section]
\newtheorem{Corrolary}{Corrolary}[section]

\newcommand{\bp}{\begin{Proposition}}   \newcommand{\ep}{\end{Proposition}}
\newcommand{\bt}{\begin{Theorem}}   \newcommand{\et}{\end{Theorem}}
\newcommand{\bl}{\begin{Lemma}}     \newcommand{\el}{\end{Lemma}}
\newcommand{\bc}{\begin{Corrolary}} \newcommand{\ec}{\end{Corrolary}}

\def\a{\alpha}

\def\m{\mu}
\def\n{\nu}

\newcommand{\Eq}[1]{(\ref{#1})}

\title{\bf Stiff phases in strongly coupled gauge theories with holographic duals}
\author{Christian Ecker${}^{1,a}$, Carlos Hoyos${}^{2,b}$, Niko Jokela${}^{3,4, c}$,\\ David Rodr\'{\i}guez Fern\'andez${}^{2,d}$, and Aleksi Vuorinen${}^{3,4,e}$\\
$\ $\\
${}^1${\em Institut f\"ur Theoretische Physik, Technische Universit\"at Wien} \\
{\em Wiedner Hauptstr. 8-10, A-1040 Vienna, Austria}\\
${}^2${\em Department of Physics, Universidad de Oviedo}\\
{\em c/.~Federico Garc\'{\i}a Lorca 18, ES-33007 Oviedo, Spain}\\
${}^3${\em Department of Physics} and ${}^4${\em Helsinki Institute of Physics}\\
{\em P.O.~Box 64, FI-00014 University of Helsinki, Finland}\\
{\small ${}^a$\tt{christian.ecker@tuwien.ac.at}, ${}^v$\tt{hoyoscarlos@uniovi.es}, ${}^c$\tt{niko.jokela@helsinki.fi},}\\ 
{\small ${}^d$\tt{rodriguezferdavid@uniovi.es}, ${}^e$\tt{aleksi.vuorinen@helsinki.fi}}
}

\begin{document}

{\twocolumn[
\begin{flushright}
FPAUO-17/10 \\
HIP-2017-13/TH\\
\end{flushright}
\maketitle
  \begin{@twocolumnfalse}
  \maketitle
    \begin{abstract}
\noindent According to common lore, Equations of State of field theories with gravity duals tend to be soft, with speeds of sound either below or around the conformal value of $v_s=1/\sqrt{3}$. This has important consequences in particular for the physics of compact stars, where the detection of two solar mass neutron stars has been shown to require very stiff equations of state. In this paper, we show that no speed limit exists for holographic models at finite density, explicitly constructing examples where the speed of sound becomes arbitrarily close to that of light. This opens up the possibility of building hybrid stars that contain quark matter obeying a holographic equation of state in their cores. 
      \vspace{0.5cm}
    \end{abstract}
  \end{@twocolumnfalse}
]}
\onecolumn

\newpage

\tableofcontents
\section{Introduction}

Astrophysical observations of neutron stars with masses up to two solar masses \cite{Demorest:2010bx,Antoniadis:2013pzd} imply that the Equation of State (EoS) relating the energy density $\varepsilon$ and pressure $p$ of the matter inside the stars should be very stiff \cite{Bedaque:2014sqa}. The stiffness can be measured by the thermodynamic derivative\footnote{The symbol $s$ denotes the entropy density here.}
\be\label{eq:vs}
v_s^2=\left(\frac{\partial p}{\partial \varepsilon}\right)_s\ ,
\ee
where $v_s$ can be identified as the speed of propagation of sound waves, naturally obeying the causal bound $v_s\leq 1$. According to our current understanding, the nature of this matter ranges from a relatively dilute gas of nuclei immersed in a sea of electrons in the crust of the star to dense nuclear and superdense neutron matter deep inside the star, expected to reach at least a few times the nuclear saturation density, $n_s\approx 0.16/\text{fm}^3$, in the cores of the most massive stars. With the deconfinement transition of Quantum Chromodynamics (QCD) expected to take place around these densities, it is at the moment still unclear, whether quark matter should be present inside the stars or not.

There are a variety of nuclear matter EoSs that predict very high speeds of sound, some of them even exceeding the speed of light \cite{Glendenning}. In all of these cases, the region of validity of the approach is, however, restricted to densities below (roughly) the nuclear saturation density, so that a straightforward extrapolation of the results to the large densities met in the cores of neutron stars is likely to suffer from uncontrollable systematic uncertainties (see \cite{Hebeler:2013nza} for a discussion of this topic). In particular, there is no hope of extending the description of these nuclear matter models to the quark matter phase, possibly relevant for the description of the stellar cores. At the same time, it is equally clear that approaches based on weak coupling expansions in the quark matter phase, such as perturbative QCD \cite{Freedman:1976ub,Vuorinen:2003fs,Kurkela:2009gj,Kurkela:2016was}, cannot be used to describe the transition region, and therefore the standard approaches for the description of this regime typically include model calculations (see e.g.~\cite{Buballa:2003qv} and references therein) and interpolations between the low- and high-density regimes \cite{Kurkela:2014vha}.

Considering the above difficulties, there is clearly room for alternative approaches to describing dense strongly interacting nuclear and quark matter. Such a novel approach could be provided by the gauge/gravity, or holographic, duality \cite{Maldacena:1997re,Gubser:1998bc,Witten:1998qj}, which offers a way to relate problems in strongly coupled field theories in their large-$N_c$ limit to calculations performed in classical supergravity in a curved spacetime. An interesting observation pointing towards neutron star matter indeed behaving like a strongly coupled system can be seen from the so-called Taub inequality \cite{Taub:1948} (see also \cite{RezzollaZanotti}),\footnote{We thank Luciano Rezzolla for drawing our attention to this inequality.} which states that in a relativistic kinetic theory causality imposes the condition
\be
\varepsilon(\varepsilon-3p)\geq \rho^2\ ,
\ee
where $\rho$ stands for the mass density.  For instance, it is easy to check that degenerate fermionic matter satisfies Taub's inequality for any value of the chemical potential. The inequality clearly implies that $\varepsilon \geq 3p$, which is saturated by conformal theories. As shown in \cite{Bedaque:2014sqa}, such an EoS is, however, too soft to support the heaviest observed stars, which clearly implies that one of the assumptions behind Taub's inequality must fail. The most likely culprit is the assumption of the validity of a quasiparticle description, which is far from being guaranteed for the matter found inside neutron stars. In fact, it may well be that the correct expansion point would be that of infinite (or very strong) coupling instead of a system of weakly coupled quasiparticles.

The holographic approach has already been used to describe both the confined \cite{Bergman:2007wp,Rozali:2007rx,Kim:2007vd,Kim:2011da,Kaplunovsky:2012gb,Ghoroku:2013gja,Li:2015uea,Elliot-Ripley:2016uwb} and deconfined \cite{Burikham:2010sw,Kim:2014pva,Hoyos:2016zke,Hoyos:2016cob} phases of QCD matter through the study of strongly coupled non-Abelian gauge field theories containing fundamental matter with a global U(1) baryon symmetry. In \cite{Hoyos:2016zke}, we adopted the strategy of describing the low-density phase of QCD matter using the Chiral Effective Theory (CET) results of \cite{Tews:2012fj}, supplemented by the extrapolations provided in \cite{Hebeler:2013nza}, and matching them with the EoS of $\mathcal{N}=2$ Super Yang-Mills theory at finite baryon density, corresponding to a D3-D7 brane intersection on the gravity side. While successful in providing a consistent description of dense QCD matter, this setup led to the prediction that the deconfinement transition would always be of such a strong first order type that the resulting hybrid stars become unstable as soon as even a microscopic amount of quark matter is generated in their cores. The reason for this behavior was found to be the soft nature of the holographic EoS, with $v_s^2<1/3$, in comparison with the stiff low-density EoSs of \cite{Hebeler:2013nza}.

The softness of the holographic EoS constructed in \cite{Hoyos:2016zke} came as no surprise; in fact, already in \cite{Hohler:2009tv,Cherman:2009tw} it was conjectured that any field theory with a gauge/gravity dual can have a speed of sound at most as large as that of a conformal theory, i.e.~$v_s\leq 1/\sqrt{3}$. In \cite{Hoyos:2016cob}, we, however, showed that this conjecture is generically not valid at finite density (even though it might hold in certain theories \cite{Cherman:2009tw}), and more recently a violation of the bound has been proposed even at zero density through the introduction of multitrace deformations in the dual gauge theory \cite{Anabalon:2017eri}. However, in both cases the violation is not nearly large enough to allow for the existence of quark matter inside neutron stars, and the question remains, whether at least a moderate softness of the EoS of strongly coupled deconfined matter is a universal prediction of holography. We should also note that a bound on the speed of sound {\em at fixed chemical potential} has been proposed in \cite{Yang:2017oer}, and it seems to hold in holographic models that reproduce thermodynamic properties of QCD computed using lattice techniques at small densities \cite{Rougemont:2017tlu}.

In the present work, we shall demonstrate that the speeds of sound obtained in gauge/gravity models can be arbitrarily close to the speed of light by considering several examples where this turns out to be the case. On the gravity side, the models consist of Einstein-Maxwell theory minimally coupled to a scalar field, which can be either charged or neutral. These models are dual to a strongly coupled gauge theory in its large-$N_c$ limit. The bulk gauge field is then dual to a global $U(1)$ current on the field theory side, while the scalar field is dual to a relevant scalar operator. A relevant deformation breaking conformal invariance is introduced by turning on a coupling for the scalar operator. The first example we will study has a string theory (top-down) realization with a known field theory dual, while the rest of the cases considered form a family of bottom-up models. Interestingly, we observe that the simplest scenario including a quadratic potential for a canonically normalized scalar field does not lead to large enough values for the speed of sound. To reach higher values, it is necessary for the scalar field to possess self-interactions, which will be reflected in the properties of higher order correlators of the dual operator. This point should be a very interesting one to investigate further in the future.

Our paper is organized as follows. In Sec.~\ref{sec:sectwo} we introduce both the top-down and bottom-up models we work with, and in Sec.~\ref{sec:secthree} we discuss a subtle issue related to the spontaneous generation of a scale in the top-down model. After this, we move on to presenting our main result, the EoS in both types of models, in Sec.~\ref{sec:secfour}, which is followed by a thorough analysis of the stability of our solutions in Sec.~\ref{sec:stability}. Conclusions are finally drawn in Sec.~\ref{sec:conclusions}, while a number of computational details will be discussed in the Appendices of the paper.

\section{Holographic models}\label{sec:sectwo}

We will use holographic models as a tool to study the EoS of strongly coupled gauge theories at finite density and temperature, although we will be more interested in low temperatures. The models will be chosen in such a way that the theory is well defined in the UV, in the sense that there is a fixed point at asymptotically large energies. If the theory was conformal, the EoS would be fixed by symmetry; here, this will be avoided by introducing a relevant deformation of the UV fixed point that breaks conformal invariance explicitly. We will consider two cases in parallel: a top-down model with a well defined string theory construction, and a family of phenomenological bottom-up models that allow a wider analysis while keeping the main ingredients of the top-down model.

\subsection{Top-down model}

The first case we are going to consider is a deformation of $\cN=4$ $SU(N_c)$ super Yang-Mills (SYM). The theory has a global $SU(4)_R\simeq SO(6)_R$ $R$-symmetry group associated to rotations of the supercharges. $\cN=4$ SYM contains vector bosons, fermions, and scalars, all in the adjoint representation of the $SU(N_c)$ gauge group. They can be listed as
\begin{center}
\begin{tabular}{ccc}
fields & symbol & $SU(4)_R$ representation\\ \hline
vector gauge bosons & $\cA_\mu$ & singlet\\
gauginos (fermions) & $\lambda^a$ & {\bf 4}\\
scalars & $\phi^I$ & {\bf 6}
\end{tabular}
\end{center}
There are three mutually commuting $U(1)_{i=1,2,3}\subset SU(4)_R$ in the $R$-symmetry group. We will study states with charge for the diagonal $U(1)$ (equal charges for all of the $U(1)_i$). Since $\cN=4$ SYM is a conformal field theory, we will also need to turn on additional couplings that break explicitly conformal invariance. We will do this by introducing a mass for the gauginos, i.e. we will add a term to the Lagrangian of the form
\be
\cL=\cL_{\cN=4}+m_0 \tr \lambda\lambda\ .
\ee
As we are not adding similar mass terms for the scalars, this also breaks supersymmetry explicitly.

In the $N_c\to \infty$ limit and for very strong 't Hooft coupling $\lambda_{YM}\gg 1$, the $\cN=4$ SYM theory has a holographic dual description as type IIB string theory in a $AdS_5\times S^5$ geometry, at weak string coupling $g_s\sim 1/N_c$ and large curvature radius compared to the string scale $L^4/(\alpha')^2\sim \lambda_{YM}$. The leading order behavior of the theory is thus captured by classical supergravity (SUGRA) in $AdS_5\times S^5$ \cite{Maldacena:1997re}. Turning on a charge density and/or additional couplings in $\cN=4$ SYM is realized in the holographic dual by turning on dual fields that modify the background geometry.

Rather than dealing with the full ten-dimensional SUGRA description of the theory, we will restrict to a subsector that admits a consistent truncation to a simpler five-dimensional theory. The truncation is explained in more detail in \cite{Hoyos:2016cob}. The action reduces to the one of Einstein-Maxwell theory coupled to two real scalars
\begin{equation} \label{eq:actiontd}
e^{-1}\cL =\frac{1}{4}R-\frac{1}{g^2}F_{\mu\nu} F^{\mu\nu}+ \frac{1}{4}(\partial_\mu\phi)^2+\frac{1}{2} \sinh^2\left(\frac{\phi}{\sqrt{2}}\right)\left(\partial_\mu \theta- 2A_\mu\right)^2-\frac{V(\phi)}{4} \ ,
\end{equation}
where $e$ is the volume density and
\begin{equation}
V(\phi)=-\frac{3g^2}{4}\left(3+\cosh(\sqrt{2}\phi)\right)\ ,
\end{equation}
with the coupling constant $g$ related to the AdS radius as $g=2/L$.  The bulk gauge field $A_\mu$ is dual to the diagonal $U(1)$ $R$-current $J^\mu$ and sources for the current. 

We introduce the complex field,
\be 
\Phi = \tanh\left(\frac{\phi}{2\sqrt{2}}\right) e^{i\theta}\ ,
\ee
in such a way that the action takes the form 
\be \label{eq:langdens}
e^{-1} \mathcal{L} = \frac{1}{4}\left[  R -L^2 F^2 -\mathcal{K}(\Phi) \vert D\Phi \vert^2 -\mathcal{V}(\Phi) \right] , \quad D_\mu \Phi= (\partial_\mu -i q A_\mu)\Phi\ ,
\ee 
with a charge $q=2$ and kinetic and potential terms
\be \label{eq:topdownL}
\mathcal{K}(\Phi) = \frac{8}{\left(  1- \vert \Phi \vert^2 \right)^2}, \quad \mathcal{V}(\Phi) = -\frac{12}{L^2}\frac{1+ \vert \Phi \vert^4}{\left(1- \vert \Phi \vert^2\right)^2}  \ .
\ee 
For small $\Phi$,
\be
\mathcal{K}(\Phi) \simeq 8, \ \ \mathcal{V}(\Phi) \simeq  -\frac{12}{L^2}\left(1+2\vert \Phi \vert^2\right)\ .
\ee
Therefore, the canonically normalized scalar has a mass $m^2 L^2=-3$ which corresponds to a field dual to an operator of conformal dimension $\Delta=3$, the gaugino mass operator $\cO=\tr\lambda\lambda$, and the associated coupling $m_0$. Therefore, the five-dimensional action of the truncated SUGRA subsector contains all the necessary ingredients for our analysis.

\subsection{Bottom-up models}

Taking the top-down model as a guide, we are going to consider a family of models with a gravity dual consisting of Einstein-Maxwell theory minimally coupled to a scalar. Thereby, we will be describing a subsector of the dual field theory including a global $U(1)$ current $J^\mu$ and a relevant scalar operator $\cO$. The usual large-$N_c$ and strong coupling limits are assumed to hold for the classical gravity approximation we take to be valid.

In order to obtain different EoSs, we will allow for some freedom in the choice of the action for the scalar field. This means that in most cases the field theory dual, if it exists, is not known. We will use these models as an exploratory mean to determine whether holographic models can produce a stiff EoS, with the perspective of looking for proper holographic duals with similar properties in the future. One can in principle allow the kinetic term and the potential for the scalar to be generic functionals, although we will fix their form to be able to do explicit calculations. The five-dimensional action for these models will be as given in \eqref{eq:langdens}. For the bottom-up models we will take the charge to be zero $q=0$, as eventually we would like to identify the $U(1)$ symmetry with baryon symmetry, which is unbroken. For simplicity, we will fix the kinetic term to be canonically normalized $\mathcal{K}(\Phi) =1$ and the potential to be of the form
\be
\mathcal{V}(\Phi) = -\frac{12}{L^2}+m^2 \vert\Phi\vert^2+\frac{V_4}{2L^2} \left(\vert\Phi\vert^2\right)^2\ .
\ee
Here, we will allow the masses to lie in the interval $0>m^2 L^2\geq -3$, in such a way the scalar field will be dual to a scalar operator of dimension in the interval $4>\Delta\geq 3$. We will study first the case with a purely quadratic potential $V_4=0$ and then the behavior when $V_4$ is changed. 

\subsection{Charged black hole solutions in the top-down model}

We take an Ansatz for the metric of the form
\begin{equation} \label{eq:holomet}
ds^2=L^2\frac{dr^2}{r^2 f(r)}+\frac{r^2}{L^2} e^{2A}\left[ -f(r)dt^2+dx_1^2 +dx_2^2+dx_3^2\right]\ ,
\end{equation}
in such a way that it is asymptotically $AdS$ at $r\to \infty$. There is a black hole horizon at $r=r_H$, where $f(r_H)=0$. The scalar field and the time component of the gauge field are also turned on and depend only on the radial coordinate, i.e.~$\Phi_0=\Phi_0(r)$, $A_0=A_0(r)$.

The equations of motion and the near boundary behavior of the bulk fields are detailed in Appendix~\ref{app:eoms}. If the fields were decoupled, their expansion at the boundary would take the form\footnote{We use a notation where $X_{(n,m)}$ is the coefficient of $(L^2/r)^m(\log (r/L))^n$ in the expansion of the field $X$.}
\be \label{eq:boundexp}
A_0 \sim \mu + \frac{L^4}{r^2}A_0{}_{(0,2)} , \quad f \sim 1 + \frac{L^8}{r^4}f_{(0,4)} , \quad A \sim 0 , \quad \Phi_0 \sim \frac{L^2}{r}\phi_{(0,1)} + \frac{L^6}{r^3}\left[ \phi_{(1,3)}\log \left(\frac{r}{L}\right) + \phi_{(0,3)}\right]\ .
\ee 
We can identify $\mu$ with the chemical potential in the dual field theory and $\phi_{(0,1)}$ with the coupling of the dual operator. If it is nonzero, this amounts to introducing a relevant deformation that breaks explicitly conformal invariance in the dual field theory. In this case, $\phi_{(0,1)}$  gives a mass to the gauginos. For $\Delta=3$, $\phi_{(0,1)}$ has dimension one, so we can in fact identify it with a mass scale $m_0\equiv \phi_{(0,1)}$. The coefficients $A_0{}_{(0,2)}$, $f_{(0,4)}$, and $ \phi_{(0,3)}$ determine the $R$-charge density, energy, and the expectation value of the scalar operator (gaugino bilinear), respectively. The coefficient of the logarithmic term $\phi_{(1,3)}$ is finally proportional to $m_0$. 

For convenience when obtaining the numerical solutions, we will perform the variable and gauge field redefinitions
\be \label{eq:ucoord}
u = \left(\frac{r_H}{r}\right)^2, \quad A_0 \to A_0 \frac{r_H}{L^2}\ ,
\ee
so that the $AdS$ boundary is now at $u\to 0$ while the horizon is at $u\to 1$. This change implies that in order to correctly match with the holographic renormalization scheme adopted, carried out in the $r$ coordinate, one must perform the shift
\be
\begin{split}  
 f_{(0,4)} &\to f_{(0,4)}+ 16 \mu ^2 m_0^2 \log\left(\frac{r_H}{L}\right) \\
  \phi_{(0,3)} &\to \phi_{(0,3)} - \left(\frac{4}{3}m_0^2 + 2 \mu ^2
\right) m_0 \log \left(\frac{r_H}{L}\right)  \\
A_0{} _{(0,2)} & \to A_0{}_{(0,2)}+8 \mu   m_0^2 \log \left(\frac{r_H}{L}\right)
\end{split} 
\ee
in such a way that the dependence on $r_H$ in the near-boundary series solution is absorbed. Such a shift has to be applied also to the boundary operators \Eq{eq:emtvevs}. In the $u$ coordinate, the near-boundary fields read
\be \label{eq:boundexpu}
\begin{split} 
\Phi_0 \sim \alpha u^{1/2} + \left[\widehat{\beta} \log (u) + \beta \right] u^{3/2}, \quad f \sim 1 + \widehat{f}_{(0)} u^2 , \quad A \sim 0 , \quad A_0 \sim \frac{r_H}{L^2}\left(a_0 + a_1 u\right)\, .
\end{split} 
\ee 
The map between the coefficients in both coordinates is
\be \label{eq:nummap}
m_0 = \frac{r_H}{L^2} \alpha	, \quad \phi_{(0,3)} = \frac{r_H^3}{L^6} \beta, \quad \mu = \frac{r_H}{L^2} a_0 , \quad A_0{}_{(0,2)} = \frac{r_H^3}{ L^6} a_1, \quad f_{(0,4)} = \frac{r_H^4}{L^8} \widehat{f}_{(0)} \ .
\ee 
Near the horizon we will impose regularity of the solution plus vanishing boundary conditions for the warp factor and the gauge field. Then, at leading order, we have
\be \label{eq:horexp}
A_0 \sim A_0{}_H^{(1)} (1-u), \quad f \sim f_H^{(1)}(1-u), \quad A \sim A_H^{(0)} + A_H^{(1)}(1-u) , \quad \Phi\sim \phi_H^{(0)}\ ,
\ee 
where the subleading terms can be found in Appendix~\ref{app:nhseries}. \\

It is convenient to define our thermodynamic variables $(\mu,T)$ in units of the mass $m_0$
\be \label{eq:reduced}
\mu_r = \frac{\mu}{m_0} , \quad  t_r = \frac{T}{m_0} \ ,
\ee
so that $a_0 = \mu_r \alpha$.  We will also normalize the thermodynamic potentials and expectation values of the charge and scalar operators by the mass and a common factor  $\cN=\frac{L^3}{16\pi G_5}$, such that
\be \label{eq:reduced2}
\varepsilon_r = \frac{\varepsilon}{\cN m_0^4} , \quad p_r = \frac{p}{\cN m_0^4} , \quad v_r = \frac{\vev{\mathcal{O}}}{\cN m_0^3} , \quad n_r = \frac{n}{\cN m_0^3}\ .
\ee
After defining 
\be 
\mathcal{W}_1 = \kappa_1 -8\log\left(m_0 L \right), \qquad \mathcal{W}_2 = \kappa_2 + \frac{32}{3}\log \left(m_0 L \right)\ ,
\ee
and taking the renormalized expectation values \Eq{eq:emtvevs}, detailed in Appendix~\ref{app:holoRG}, we then get 
\be
\begin{split} \label{eq:renvevs}
\varepsilon_r & =-\frac{3 \widehat{f}_0}{\alpha ^4} -\frac{8 \beta }{\alpha ^3} +\log(\alpha ) \left(32 \mu _r^2-\frac{32}{3}\right)-4\mu_r^2\left(\kappa_1 +3 \right) -\kappa_2-\frac{16}{3} \\
p_r &= -\frac{\widehat{f}_0}{\alpha ^4}+\frac{8 \beta }{\alpha ^3} +\log (\alpha) \left(32 \mu _r^2+\frac{32}{3}\right)-4\mu_r^2\left( \kappa_1 +1\right) +\kappa_2 +\frac{16}{3} \\
v_r &= 32\frac{\beta }{\alpha ^3}+\frac{64}{3} \log (\alpha ) \left(3 \mu_r^2+2\right)-8 (\kappa_1+4)\mu_r^2+4 \kappa_2 +\frac{32}{3}\\
n_r &= -8 \left[a_1 -8 \mu_r \log (\alpha)+(\kappa_1+4) \mu_r\right]\ .
\end{split}
\ee 
We have computed the solutions by means of the shooting technique, thoroughly explained in Appendix~\ref{app:shooting}. We plot the results as a function of $\mu_r$ for a fixed temperature $t_r=1$ in Fig.~\ref{fig:topdnumsols}.

\begin{figure}[h!]
\begin{center}
\begin{tabular}{cc}
 \includegraphics[width=6cm]{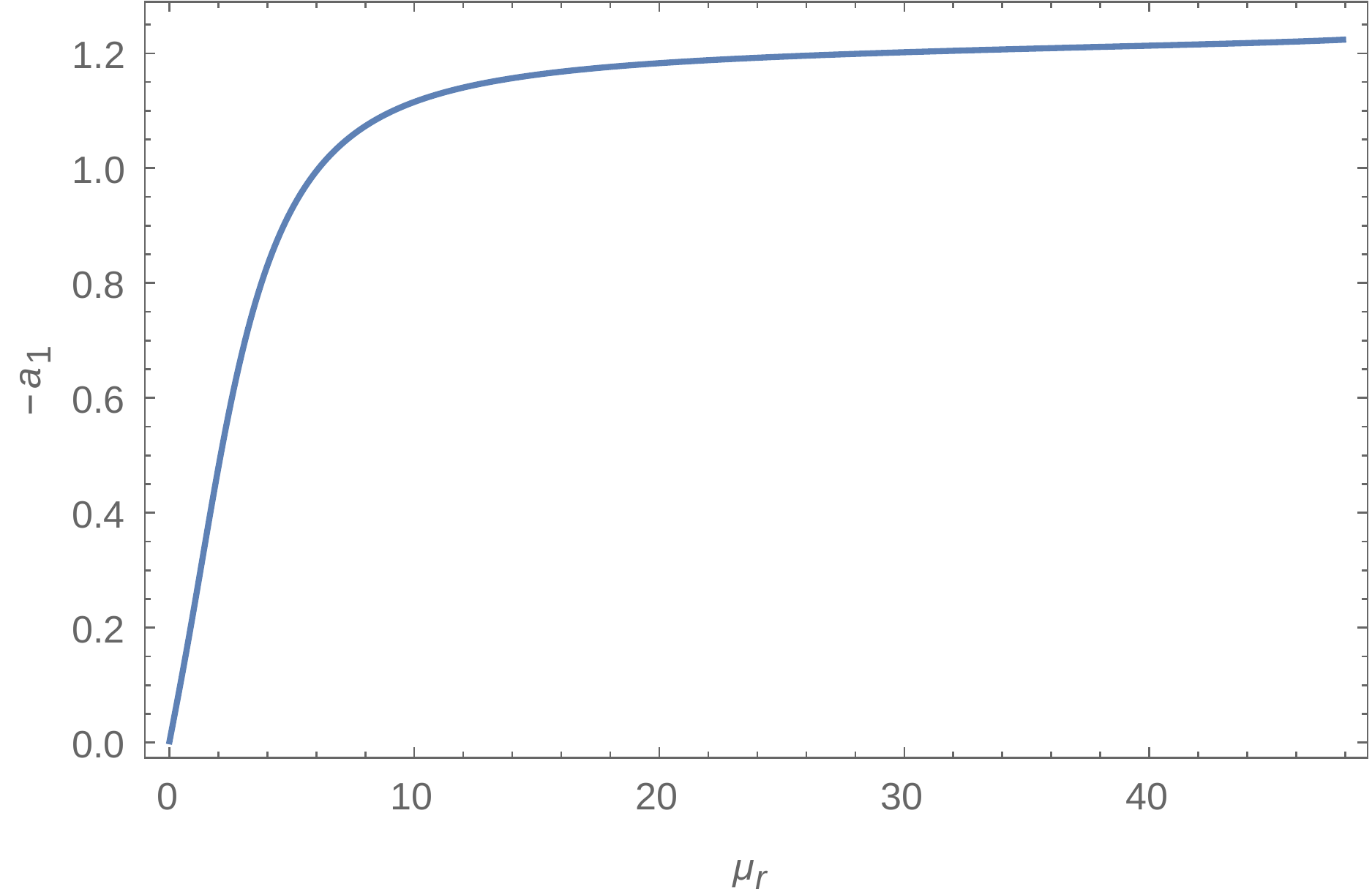} &   \includegraphics[width=6cm]{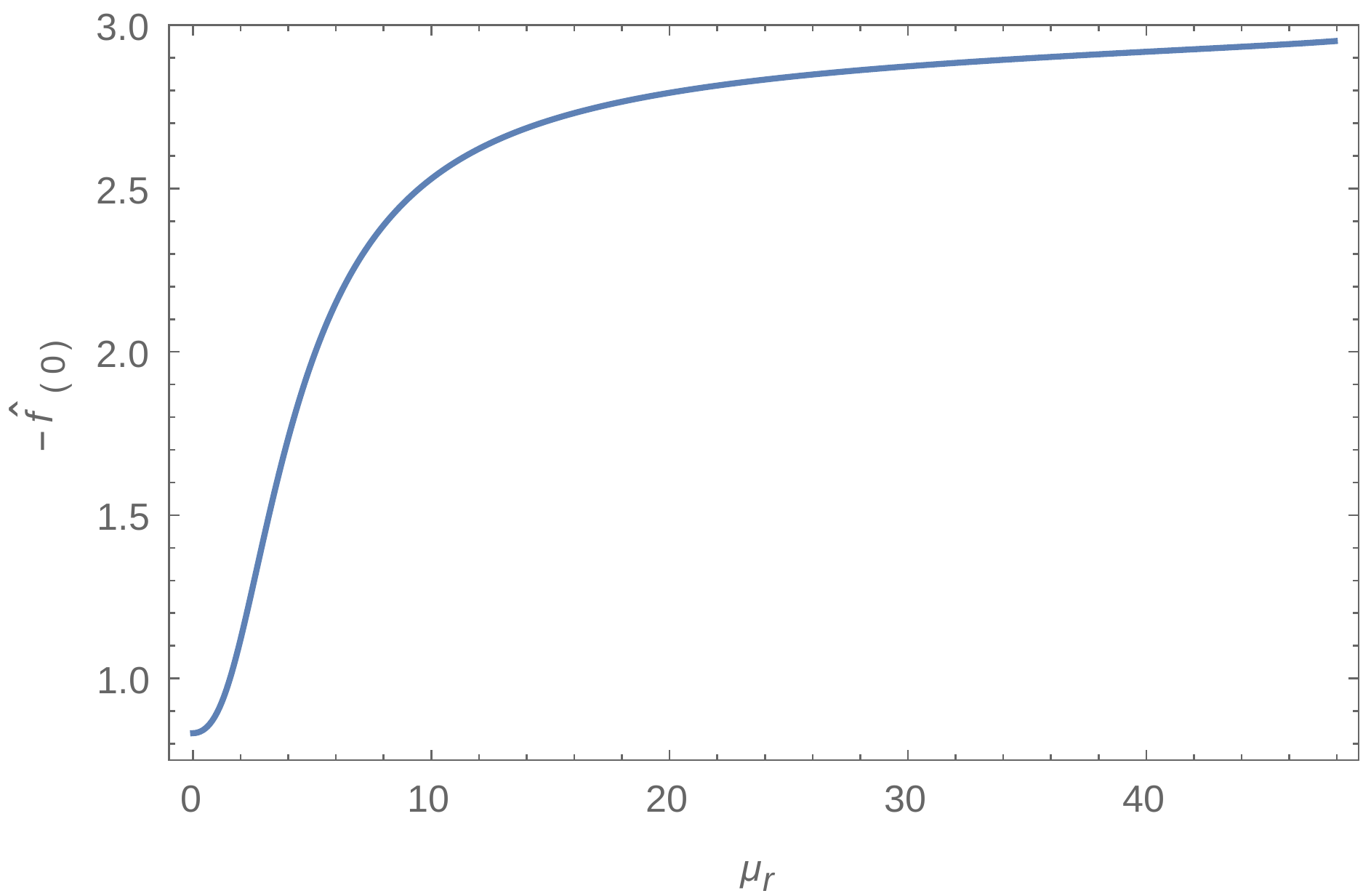} \\
  \includegraphics[width=6cm]{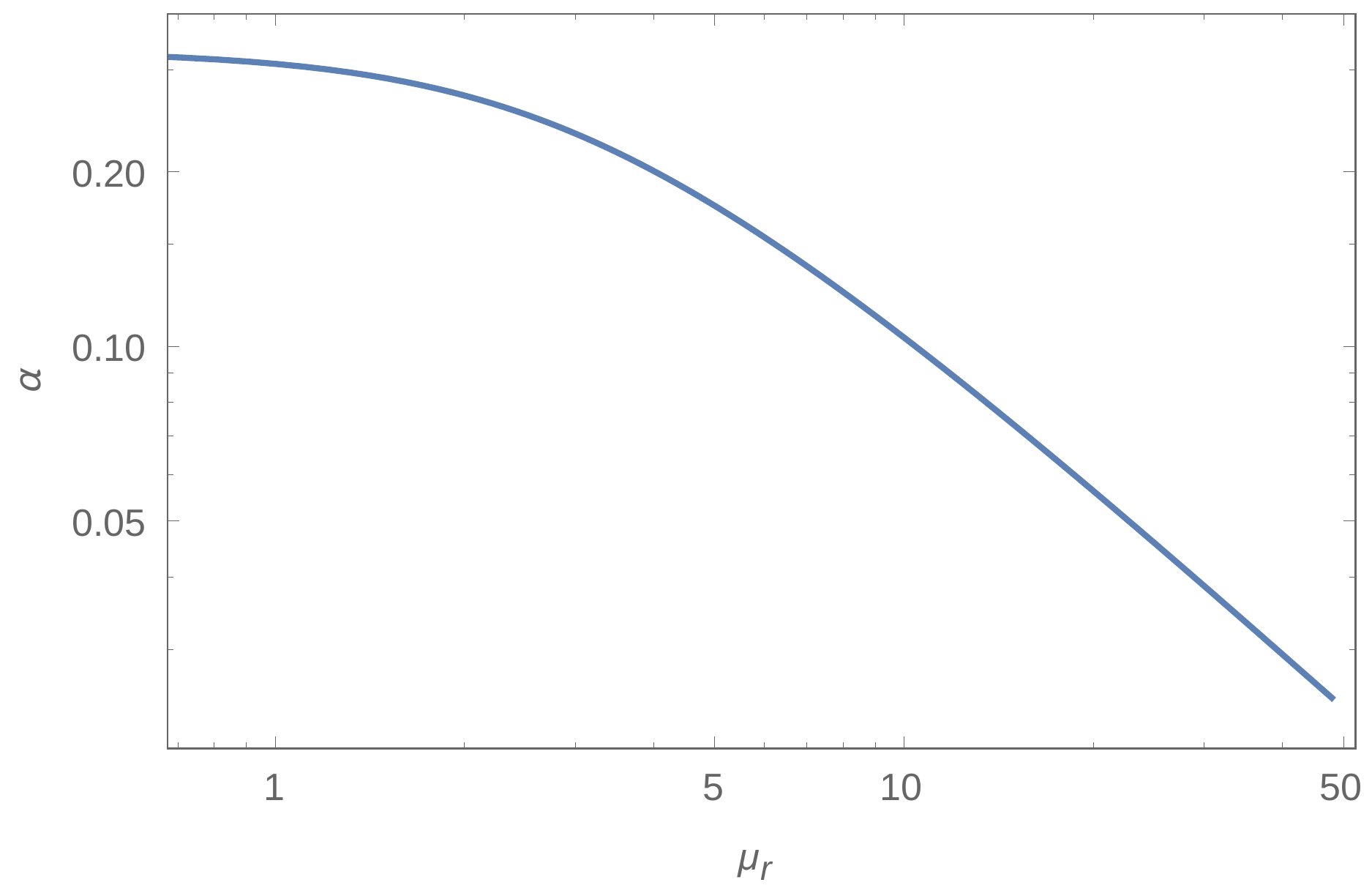} &  \includegraphics[width=6cm]{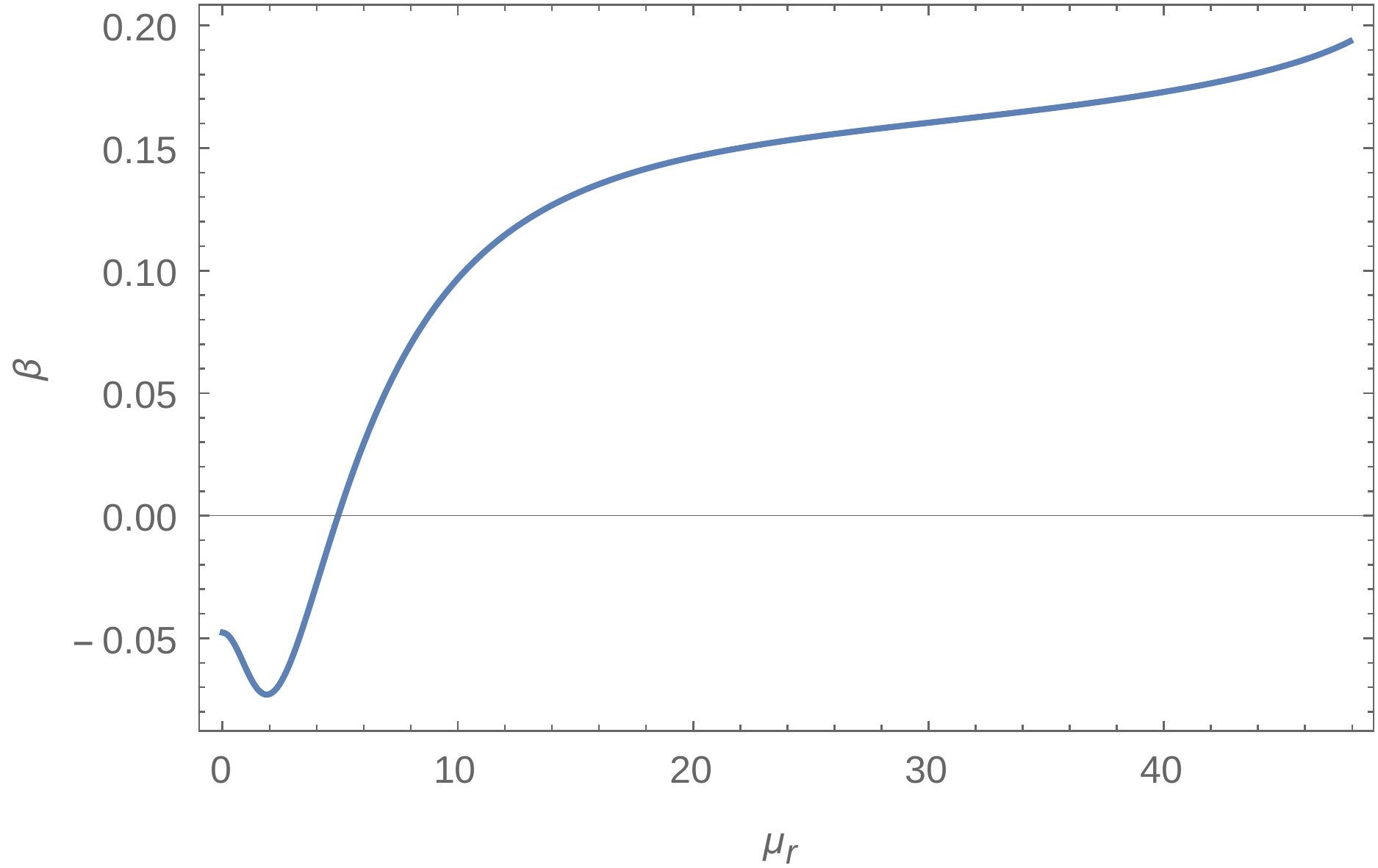}
\end{tabular}
\caption{\small Coefficients of the numerical solutions defined in Eq.~\eqref{eq:nummap} as functions of the reduced chemical potential $\mu_r$. From left to right and top to bottom, $-a_1$, $-\widehat{f}_{(0)}$, $\alpha$, and $\beta$.}\label{fig:topdnumsols}
\end{center}
\end{figure}

\subsection{Charged black hole solutions in bottom-up models}

In the bottom-up models, we will proceed in a similar manner to the top-down one. We take an Ansatz for the metric of the form given in \eqref{eq:holomet}, and fix $q=0$ for simplicity and because the potential application to the physics of dense nuclear matter requires the $U(1)$ symmetry to be unbroken. The equations of motion and the near boundary behavior of the bulk fields are detailed in the appendices of \cite{Hoyos:2016cob}.  For $\Delta=3$ the equations and expansions take a similar form as in the top-down model \eqref{eq:boundexp}. For $\Delta\neq 3$ only the expansion of the scalar field at the boundary changes to
\be 
 \Phi_0 \sim \frac{L^{2(4-\Delta)}}{r^{4-\Delta}}\tilde{\phi}_{(0,0)} + \frac{L^{2\Delta}}{r^\Delta} \phi_{(0,0)} \ .
\ee 
We can identify $\tilde{\phi}_{(0,0)}$ with the coupling of the dual operator. If it is nonzero, this amounts to introducing a relevant deformation that breaks explicitly conformal invariance in the dual field theory. Similarly to the top-down model, we will introduce the mass scale $m_0=(\tilde{\phi}_{(0,0)})^{1/(4-\Delta)}$.

For convenience when obtaining the numerical solutions, we will perform the change to the $u$  coordinate \eqref{eq:ucoord}. The near-boundary expansions of the fields are given by Eq.~\eqref{eq:boundexpu}, except for the scalar field, which now reads
\be 
\begin{split} 
\Phi_0 \sim \alpha u^{(4-\Delta)/2} + \beta  u^{\Delta/2}\ .
\end{split} 
\ee 
The map between the coefficients in both coordinates is given by \eqref{eq:nummap}, except for the scalar, which now takes the form
\be \label{eq:nummapbot}
\tilde{\phi}_{(0)} = \left(\frac{r_H}{L^2}\right)^{4-\Delta} \alpha	, \quad \phi_{(0)} = \left(\frac{r_H}{L^2}\right)^\Delta \beta\ .
\ee 
Near the horizon we will impose regularity of the solution plus vanishing boundary conditions for the warp factor and the gauge field as in \eqref{eq:horexp}. 

It is convenient to define our thermodynamic variables $(\mu,T)$ in units of the mass $m_0$, as in \eqref{eq:reduced} and \eqref{eq:reduced2}. The normalization of the expectation value of the scalar operator reads in the general case
\be 
v_r = \frac{\vev{\mathcal{O}}}{\cN m_0^\Delta} \ .
\ee
Taking the renormalized expectation values detailed in the Appendix of \cite{Hoyos:2016cob}, we get for $\Delta=3$:
\be
\begin{split} \label{eq:renvevsbot3}
\varepsilon_r & =-\frac{3 \widehat{f}_0}{\alpha ^4} -\frac{ \beta }{\alpha ^3} +\left(\frac{1}{3}+\frac{V_4}{2} \right)\log(\alpha ) -\kappa_2-\frac{1}{12}-\frac{V_4}{8} \\
p_r &= -\frac{\widehat{f}_0}{\alpha ^4}+\frac{\beta }{\alpha ^3} -\left(\frac{1}{3}+\frac{V_4}{2} \right)\log(\alpha )+\kappa_2 +\frac{1}{12}+\frac{V_4}{8}\\
v_r &= -2\frac{\beta }{\alpha ^3}+2\left(\frac{1}{3}+\frac{V_4}{2} \right)\log (\alpha )-2 \kappa_2 -\frac{1}{3}-\frac{V_4}{2}\\
n_r &= -8 a_1 \ .
\end{split}
\ee 
We will fix $\kappa_2=0$ in the following, since this parameter is irrelevant for the speed of sound. This selects $V_4=-2/3$ as a special value, for which the logarithmic terms drop and the conformal anomaly vanishes, although there are still terms contributing to the trace of the energy-momentum tensor proportional to the expectation value of the scalar operator.

\begin{figure}[h!]
\begin{center}
\begin{tabular}{cc}
 \includegraphics[width=6.5cm]{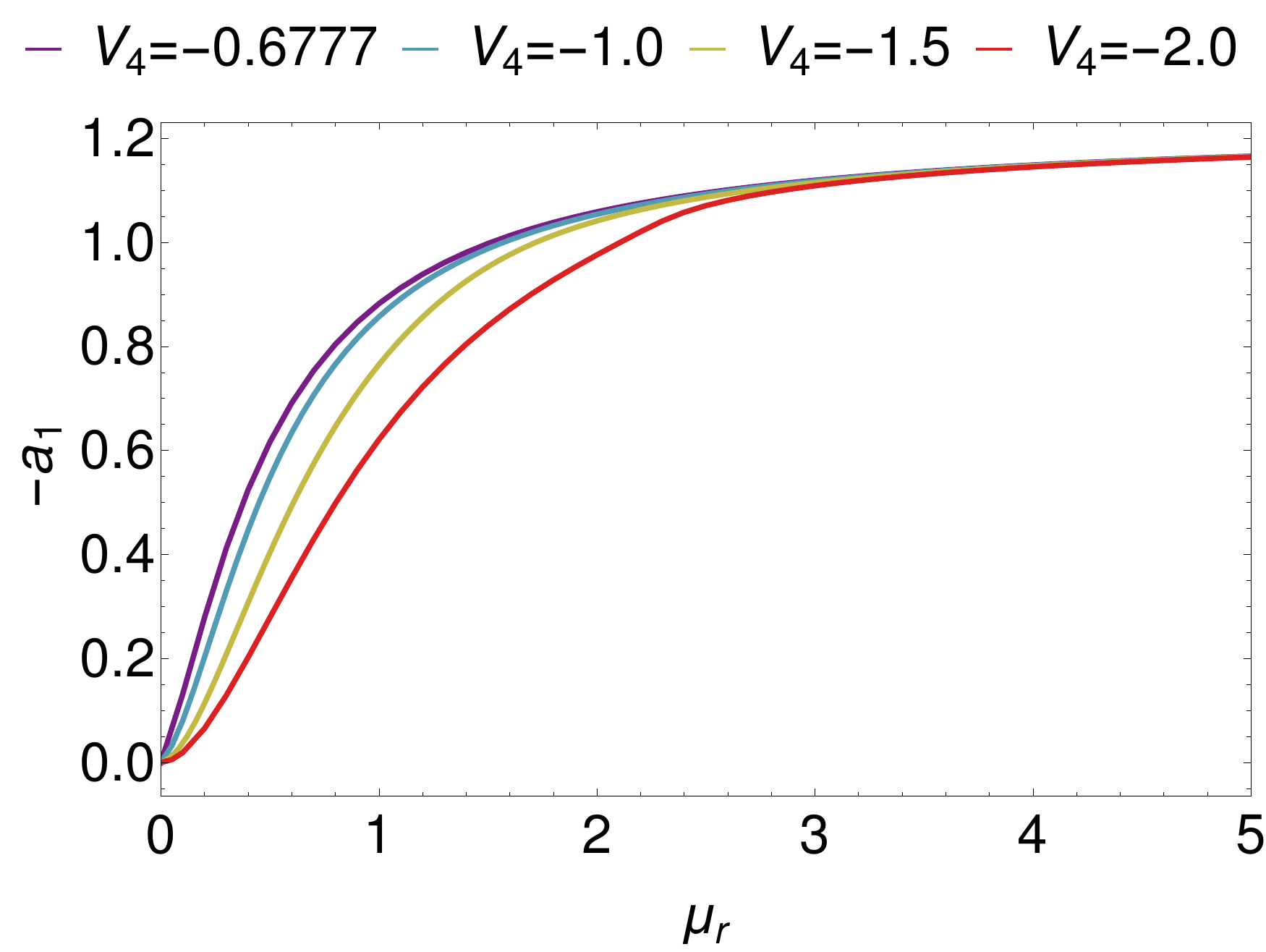} &   \includegraphics[width=6.5cm]{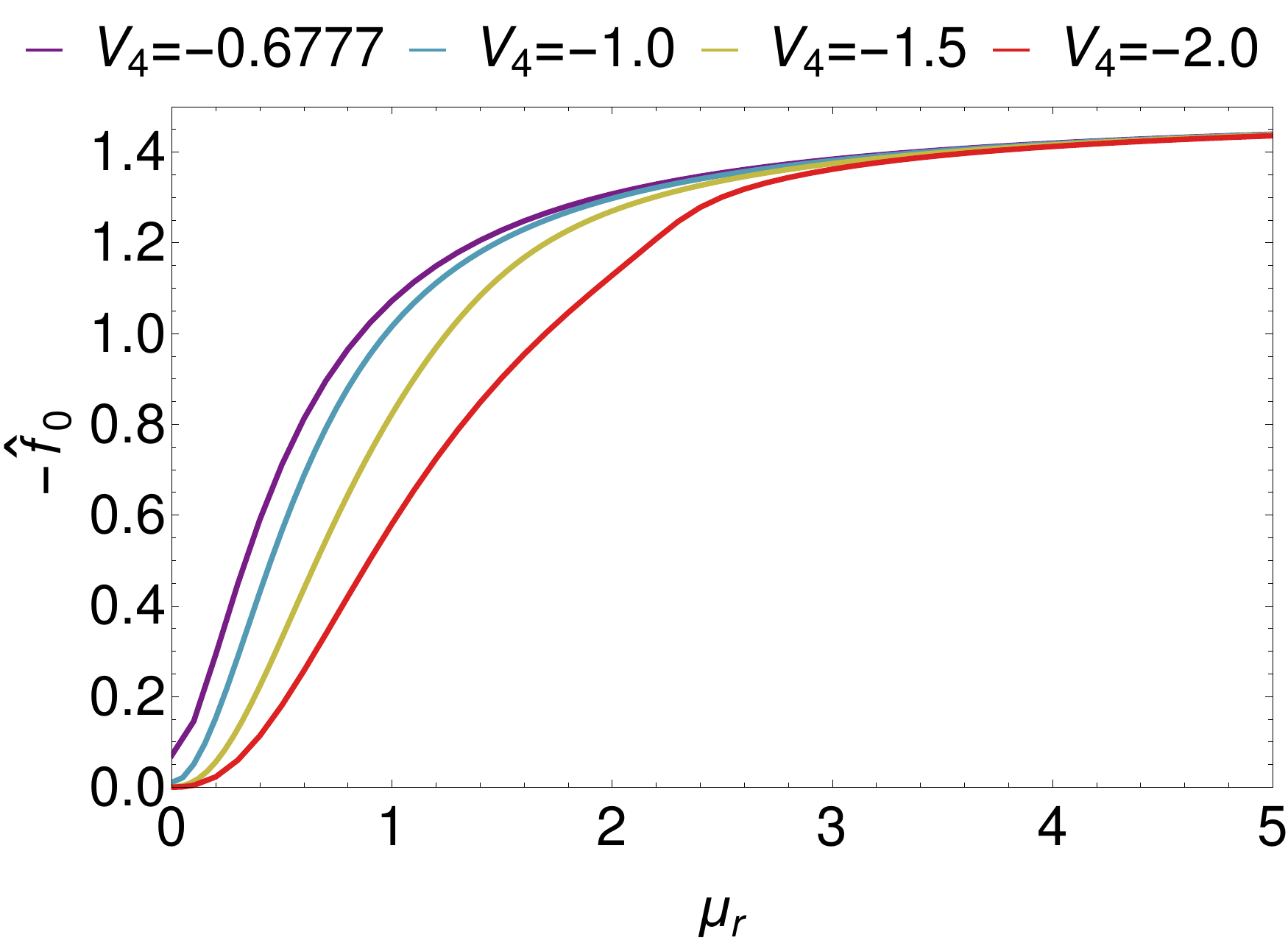} \\
  \includegraphics[width=6.5cm]{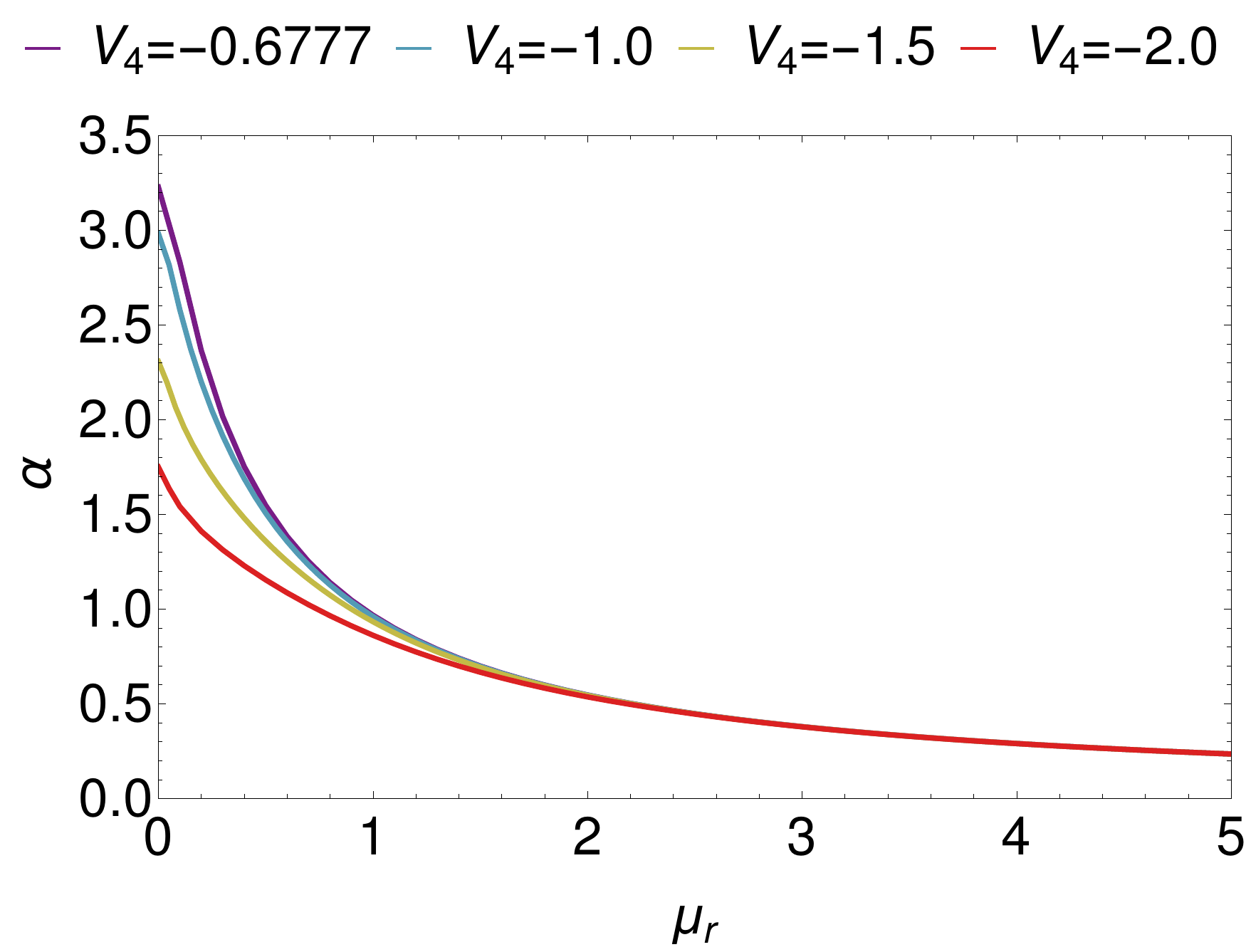} &  \includegraphics[width=6.5cm]{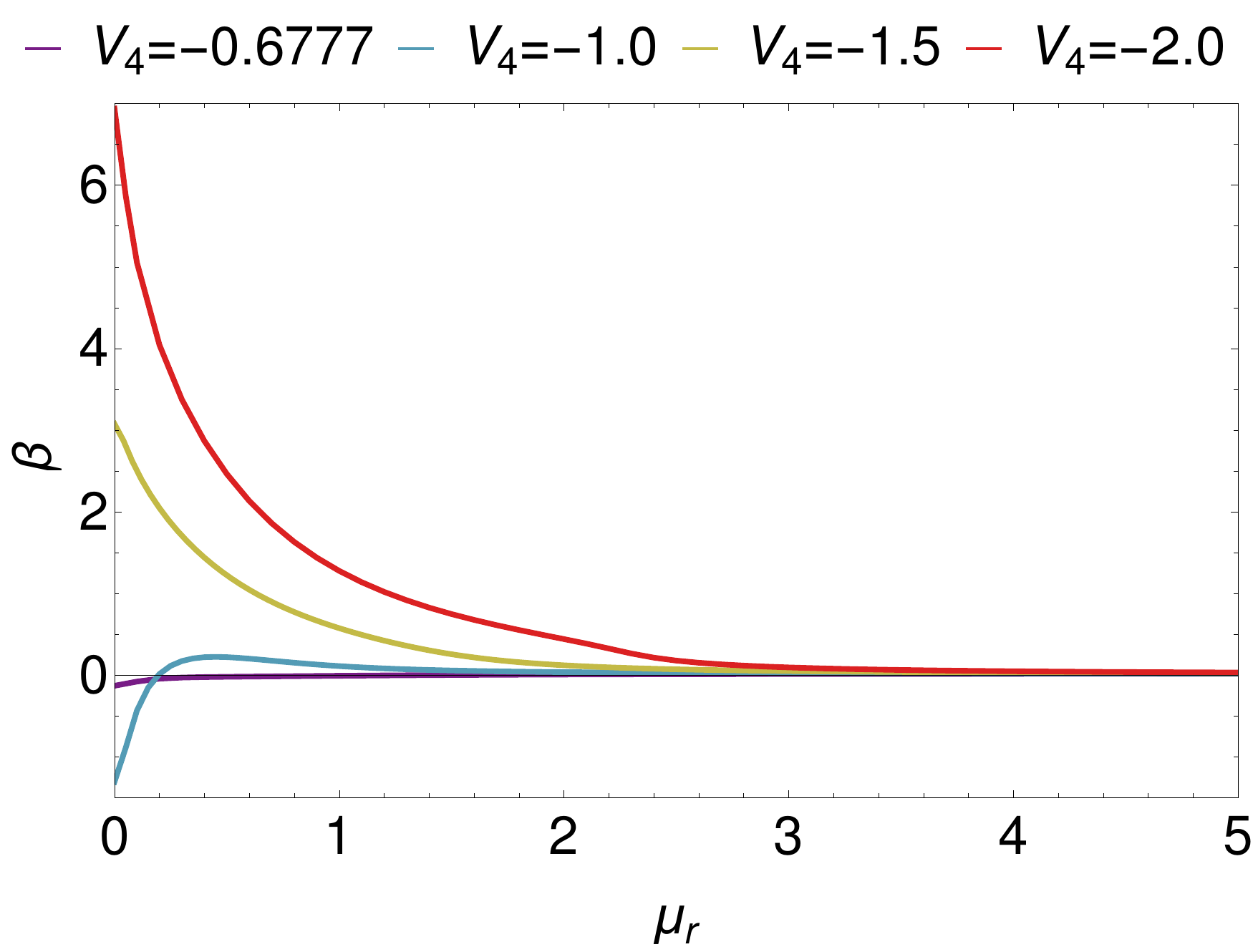}
\end{tabular}
\caption{\small Coefficients of the numerical solutions in the bottom-up case as functions of the reduced chemical potential $\mu_r$ at fixed temperature $t_r=0.1$ and different values of $V_4$. From left to right and top to bottom, $-a_1$, $-\widehat{f}_{(0)}$, $\alpha$, and $\beta$.}\label{fig:botunumsols}
\end{center}
\end{figure}

For $\Delta\neq 3$, we on the other hand get
\be
\begin{split} \label{eq:renvevsbotD}
\varepsilon_r & =-\frac{3 \widehat{f}_0}{\alpha ^4} +(\Delta-4)(\Delta-2)\frac{\beta }{\alpha ^3} \\
p_r &= -\frac{\widehat{f}_0}{\alpha ^4}-(\Delta-4)(\Delta-2)\frac{ \beta }{\alpha ^3}\\
v_r &= -2(\Delta-2)\frac{\beta }{\alpha ^3}\\
n_r &= -8 a_1 \ .
\end{split}
\ee 
We have computed the solutions using the same numerical methods as for the top-down model. The results are plotted as functions of $\mu_r$ for a fixed temperature $t_r=0.1$ in Fig.~\ref{fig:botunumsols}.

\section{Generation of a new scale in the top-down model}\label{sec:secthree}

There are some subtleties entering the EoS of the top-down model that we shall presently discuss. In \eqref{eq:renvevs}, $\kappa_1$ and $\kappa_2$ are the coefficients of finite counterterms. These terms are scheme dependent but once the renormalization scheme has been fixed, their values are related to physical quantities such as the expectation value of the scalar operator and the charge density. This implies that the theory is not completely determined by the bulk action of the gravity dual, but it is necessary to specify the value of the finite counterterms as well.

From the point of view of the field theory, consider that in addition to the $\cN=4$ SYM fields there is a decoupled scalar field $\varphi$ and a Yukawa coupling $Y_\varphi$ between the scalar and the $\cN=4$ SYM gauginos,
\be
\cL_{Y}=Y_\varphi \varphi \tr \lambda\lambda \, .
\ee
In the large-$N_c$ limit, we can treat the scalar field as quenched, neglecting loop effects from the $\cN=4$ SYM theory. Nevertheless, this coupling breaks conformal invariance (even though it is classically marginal) and will introduce a logarithmic dependence $\log(E/\Lambda)$ on the energy scale $E$ in physical observables, such as scattering cross sections. In particular, a wave function renormalization of $\varphi$ will show up in the kinetic term of the scalar field, having the same form as the finite counterterm associated to $\kappa_1$. The scale $\Lambda$ that appears inside the log depends on the scheme, but can be fixed by measurement. After this, the value of $\Lambda$ will be different in different schemes, but physical quantities will naturally have the same values in each of them.

On top of the scale appearing due to logarithmic terms, if the scalar field acquires an expectation value $\vev{\varphi}=m_0$, this will affect the $\cN=4$ SYM theory as an explicit breaking of conformal invariance. Note that in principle the scale of explicit breaking $m_0$ and the scale that determines the running of the coupling $\Lambda$ would be completely independent, if no further condition is imposed.

To illustrate the above with an example, consider the computation of a one-loop contribution to the self-energy of a scalar field due to a loop of a fermion field of mass $m_0$. There is a logarithmic UV divergence that in dimensional regularization in $d=4-2\epsilon$ dimensions becomes a pole as $\epsilon\to 0$. Depending on the scheme, removing this divergence leaves behind different finite terms, taking the forms
\be
\begin{split}
\Sigma_{MS}(p^2) &\sim \beta  m_0^2\left(-\gamma_E+\log(4\pi)+\log\frac{\sqrt{-p^2}}{\Lambda}\right)\\
\Sigma_{\overline{MS}}(p^2) &\sim \beta  m_0^2 \log\frac{\sqrt{-p^2}}{\bar{\Lambda}}\\
\Sigma_{FS}(p^2) &\sim m_0^2\left(\kappa_{FS}+\beta\log\frac{\sqrt{-p^2}}{m_0}\right)\, .
\end{split}
\ee
Here, $\beta\sim Y_\varphi^2$ is a scheme-independent factor, $MS$ and $\overline{MS}$ denote the usual (modified) minimal subtraction schemes with scale parameters $\Lambda$ and $\bar{\Lambda}$, and $FS$ stands for a fixed scale scheme with an arbitrary finite term $\kappa_{FS}$. The physical mass of the scalar $M$ corresponds to the position of the pole in the propagator
\be
p^2+\Sigma(p^2)\Big|_{p^2=-M^2}=0 \ ,
\ee
where $m_0$ is the bare mass. This can be viewed as fixing the arbitrary renormalization scales of the $MS$ and $\overline{MS}$ schemes and the constant in the $FS$ scheme,
\be
\bar{\Lambda} =\frac{e^{\gamma_E}}{4\pi}\Lambda=M e^{-M^2/\beta m_0^2},\ \ \kappa_{FS}=\frac{M^2}{m_0^2}-\beta \log\frac{M}{m_0}\ .
\ee
For a given scheme, changing the renormalization scale or the finite counterterm amounts to a change of the physical scale and thus a modification of the theory. 

In the holographic calculation we fix the scheme of holographic renormalization by using $L$ as the reference scale in the asymptotic expansion of the fields and $m_0$ in the definition of the finite counterterms. We could have chosen a different scale, say $L'$, in such a way that
\be
\mathcal{W}_1 = \kappa_1' -8\log\left(m_0 L' \right), \qquad \mathcal{W}_2 = \kappa_2' + \frac{32}{3}\log \left(m_0 L' \right)\ .
\ee
Physical results would be unchanged as long as we appropriately identify the values of the finite counterterms in each scheme,
\be
\kappa_1'=\kappa_1-8\log(L/L'), \ \ \kappa_2'=\kappa_2+\frac{32}{3}\log(L/L')\ .
\ee
We could also have changed the scheme by using a scale different from $m_0$ in the logs
\be
\mathcal{W}_1 = \kappa_1' -8\log\left(m' L \right), \qquad \mathcal{W}_2 = \kappa_2' + \frac{32}{3}\log \left(m' L \right)\ ,
\ee
leading to a somewhat different relation between the finite counterterms in different schemes,
\be
\kappa_1'=\kappa_1-8\log(m_0/m'), \ \ \kappa_2'=\kappa_2+\frac{32}{3}\log(m_0/m')\ .
\ee
This shows that an arbitrary scale can indeed be introduced through holographic renormalization.

Once we have fixed our renormalization scheme (for instance one could choose schemes where $\kappa_1'=0$ or $\kappa_2'=0$), different values of finite counterterms correspond to different values of physical quantities (i.e.~renormalization group invariants). However, one can see that the effect of $\kappa_2$ is to add a term independent of the temperature or the chemical potential that shifts the value of the vacuum energy. It is therefore unimportant for thermodynamics, and a valid physical choice could be that the effective cosmological constant term in the dual field theory vanishes. A similar term appears in the D3/D7 model \cite{Karch:2002sh}, where the counterterm is fixed by supersymmetry and gives a vanishing expectation value for the scalar operator \cite{Karch:2005ms}.

\begin{figure}[ht!]
	\begin{center}
\includegraphics[width=012cm]{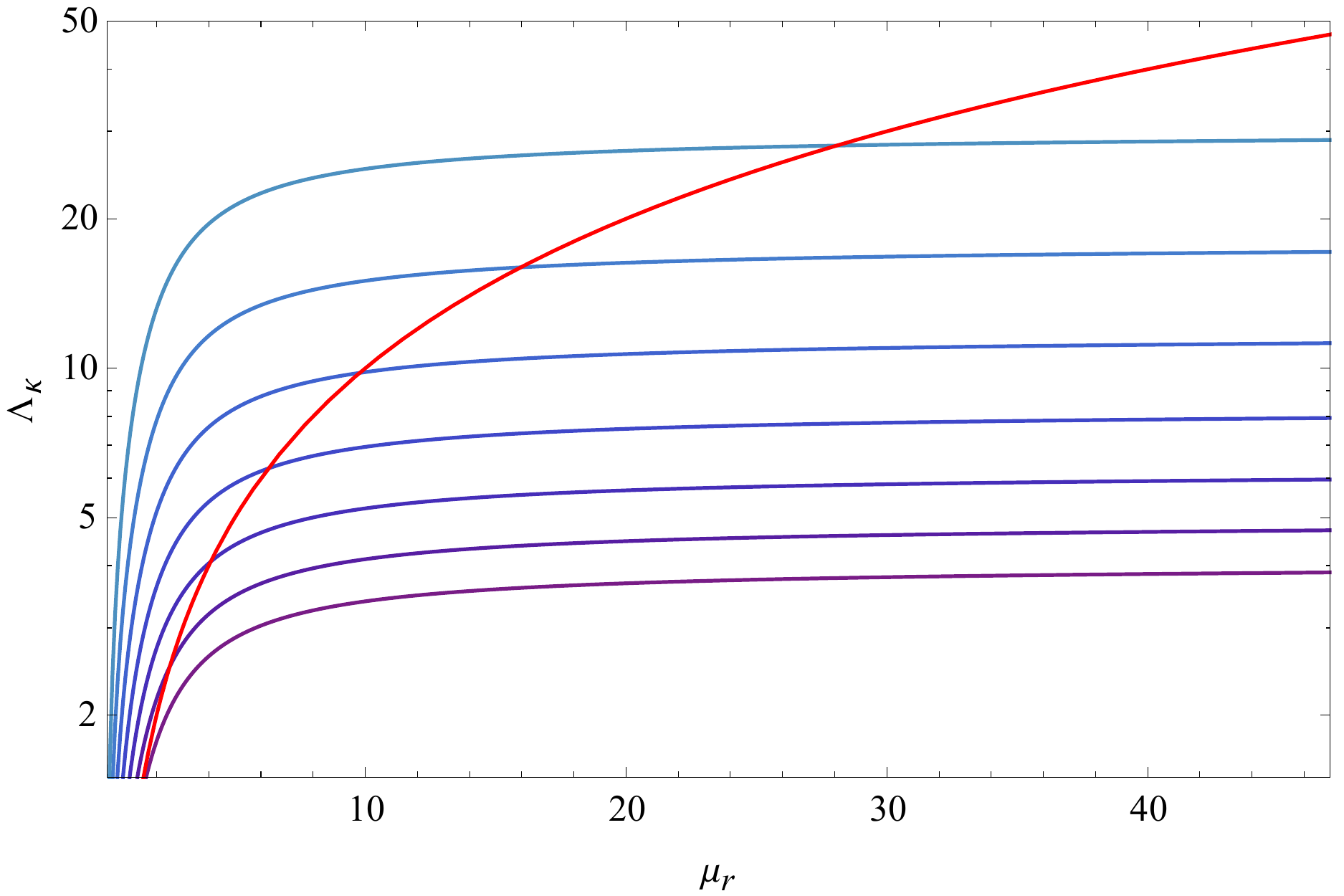}
\caption{\small $\Lambda_\kappa$ vs $\mu_r$ for different values of $\kappa_1<0$. The crossings with the red line correspond to points where $\Lambda_\kappa=\mu_r$. From bottom to top, $\kappa_1 = -9.49$, $\kappa_1=-11.05$, $\kappa_1=-12.94$, $\kappa_1=-15.22$, $\kappa_1=17.99$, $\kappa_1= -21.38$, $\kappa_1=-25.53$. } \label{fig:lambdak}	
	\end{center}
\end{figure}

Compared to $\kappa_2$, $\kappa_1$ has a more interesting and physical effect: it changes the argument of logarithms of $\alpha$ according to
\be
\log(\alpha)\longrightarrow \log\left(\alpha e^{-\kappa_1/8}\right)=\log\left(\frac{a_0 e^{-\kappa_1/8}}{\mu_r}\right)\equiv \log\left(\frac{\Lambda_\kappa}{\mu}\right)\ . \label{eq:logs}
\ee
This means that a new scale $\Lambda_\kappa$ has been spontaneously generated in the dual field theory, and that its relative size in comparison with the scale of the explicit breaking of conformal invariance is controlled by $\kappa_1$:
\be\label{eq:lambdak}
\frac{\Lambda_\kappa}{m_0}=a_0 e^{-\kappa_1/8} \ .
\ee
In particular, for $|\kappa_1|$ sufficiently large, $\Lambda_\kappa$ can be pushed towards the UV. In Fig.~\ref{fig:lambdak} we plot $\Lambda_\kappa$ as a function of the reduced chemical potential for various negative values of $\kappa_1$. When the red line crosses the other curves, $\Lambda_\kappa=\mu_r$ and the argument of the logarithm in Eq.~(\ref{eq:logs}) becomes unity.

\section{Equation of State}\label{sec:secfour}

If a weakly coupled quasiparticle description is possible for the system under study, it is appropriate to use kinetic theory to derive its Equation of State. In a relativistic theory causality then imposes a constraint, Taub's inequality  \cite{Taub:1948}
\be
\tau=\frac{\varepsilon(\varepsilon-3p)}{\rho^2}\geq 1 \ ,
\ee
where $\rho$ is the mass density. One can check for instance that for a degenerate (non-interacting) Fermi liquid $\tau=\tau_F\geq 1$, where
\be
\tau_{F}=\frac{9}{16 \left(\mu _r^2-1\right){}^3} \left[2 \mu _r^6-3 \mu _r^4+\mu _r^2+\log ^2\left(\mu _r+\sqrt{\mu
   _r^2-1}\right)-2 \sqrt{\mu _r^2-1} \mu _r^3 \log \left(\mu _r+\sqrt{\mu
   _r^2-1}\right)\right]
\ee
and $\mu_r=\mu/m_F$ where $m_F$ is the mass of the fermions.

In a strongly coupled theory the above condition may easily be violated. A simple example is the D3/D7 model \cite{Karch:2002sh} that is used to model flavor physics at strong coupling, and that contains quarks and squarks with a mass $m_q$. The EoS is known analytically  \cite{Karch:2007br,Karch:2008fa,Karch:2009eb,Ammon:2012je,Itsios:2016ffv}, and the pressure, energy density, and mass density at zero temperature read as functions of the chemical potential
\be
p=\lambda(\mu^2-m_q^2)^2,\ \ \varepsilon=\lambda(\mu^2-m_q^2)(3\mu^2+m_q^2),\ \ \rho=4\lambda m_q\mu(\mu^2-m_q^2) \ ,
\ee
where $\lambda$ is an unimportant constant factor. Defining the reduced chemical potential as $\mu_r=\mu/m_q$, one finds
\be
\tau_{D7}=\frac{3}{4}\left(1+\frac{1}{3\mu_r^2} \right) \ \Rightarrow \ 1\geq \tau_{D7}\geq \frac{3}{4} \ .
\ee
Taub's inequality is obviously violated, which indicates that the theory is indeed strongly coupled and that it possesses no good quasiparticle description. Note that there is, however, a (weaker) bound that constrains the Equation of State. Indeed, as long as $\tau\geq 0$ we will have a condition
\be
\varepsilon\geq 3p \ .
\ee

\subsection{Top-down model}

\begin{figure}[t!]
	\begin{center}
	\begin{tabular}{cc}
\includegraphics[width=6.5cm]{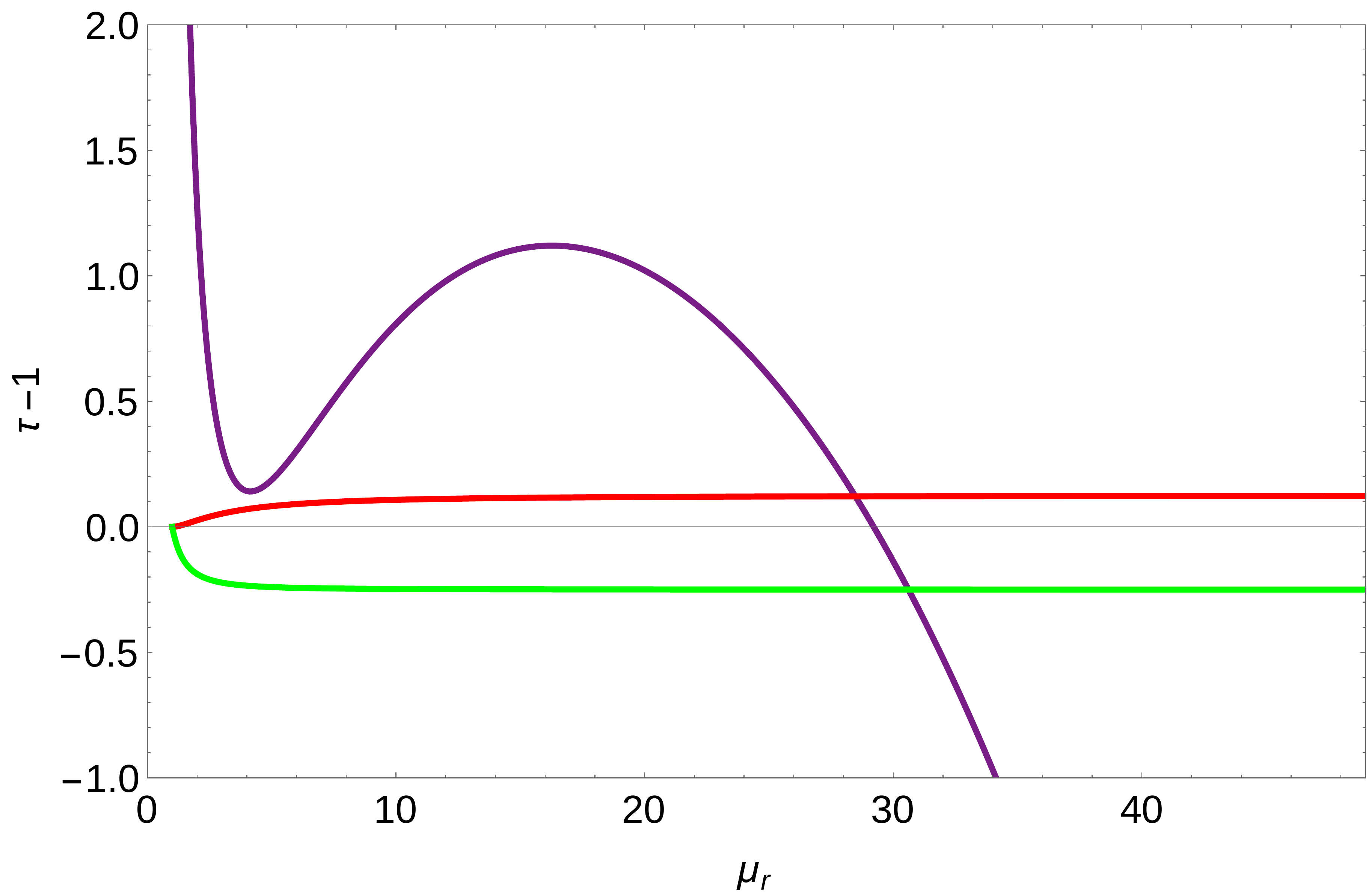} & \includegraphics[width=6.5cm]{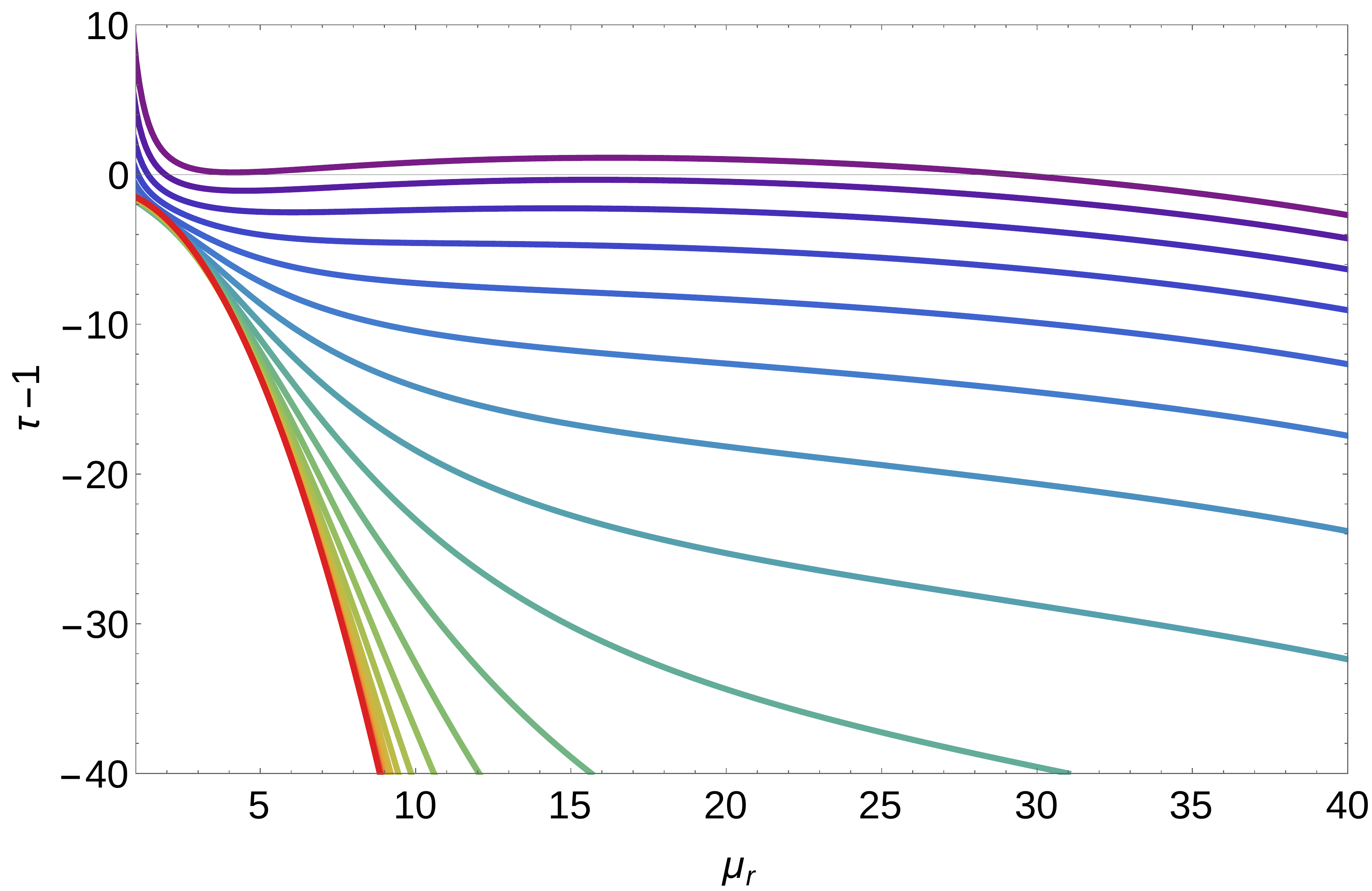}
\end{tabular}
\caption{\small Left plot: The function $\tau-1$ appearing in Taub's inequality plotted as a function of the chemical potential for a degenerate Fermi liquid (red), the D3/D7 model (green) and the top-down model for $\kappa_1=-12.86$ (purple). Holographic models clearly violate Taub's inequality $\tau\geq 1$. At large values of $\mu_r$, $\tau-1$ in the D3/D7 model approaches a negative constant corresponding to  $\tau=3/4$, while the supergravity curve keeps decreasing and reaches $\tau=0$ at $\mu_r\simeq 34$. The difference in behavior can be understood from the fact that   $\varepsilon-3p\sim m_q^2 \mu^2$ in the D3/D7 model, while $\varepsilon-3p\sim m_0^2 \mu^2\log(m_0/\mu)$ in the supergravity model. On the right plot we show Taub's inequality for the top-down model for different values of $\kappa_1$, spanning from $-12.86$ (upper curve) to $-5.18\times 10^3$ (bottom curve).}	\label{fig:Taub} 
	\end{center}
\end{figure}

\begin{figure}[t]
	\begin{center}
	\begin{tabular}{cc}
\includegraphics[width=6.5cm]{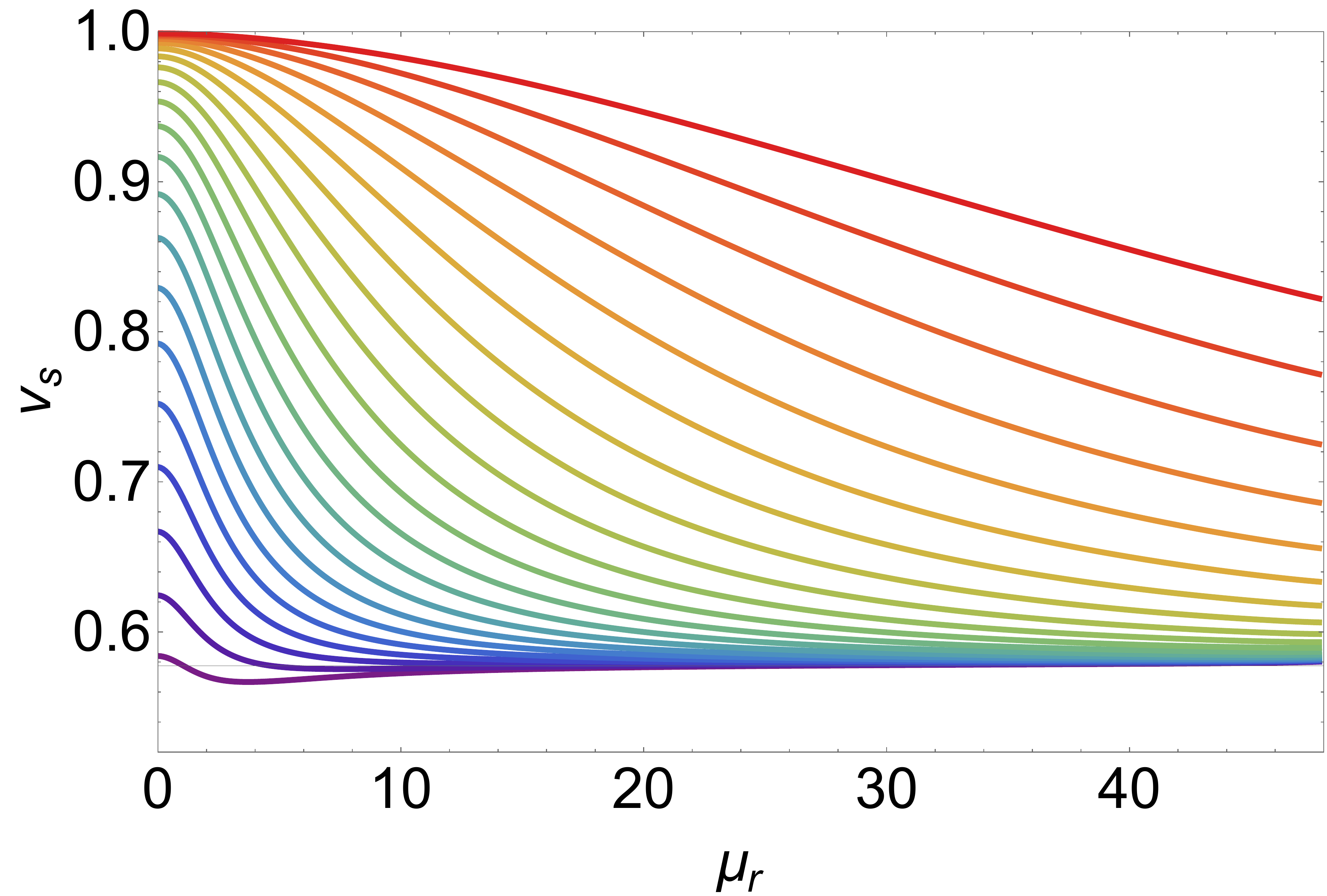} & \includegraphics[width=6.5cm]{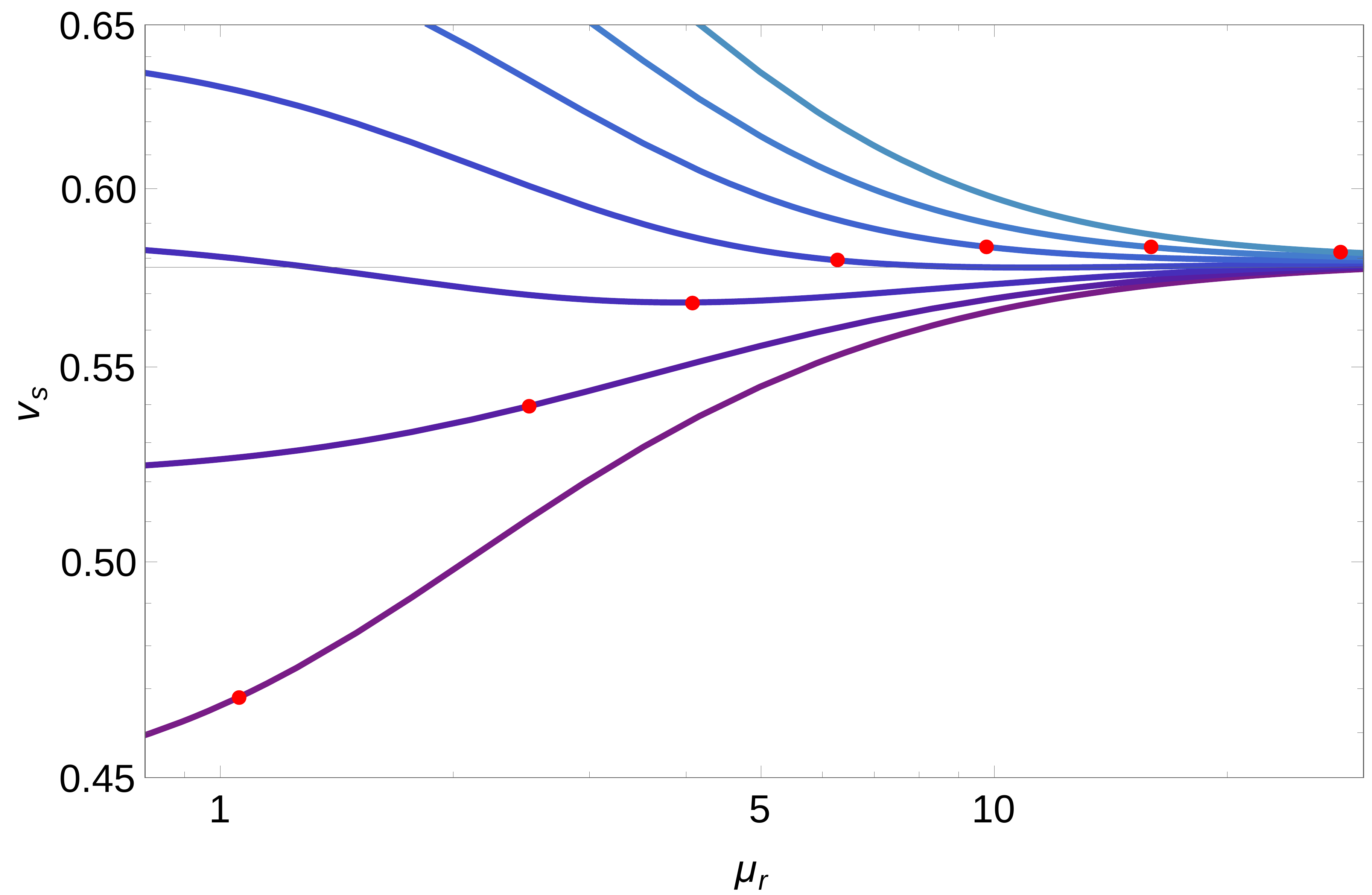}
\end{tabular}
\caption{\small $v_s$ as a function of the reduced chemical potential at $t_r=1$ at different values of $\kappa_1$. The thin horizontal line corresponds to the value of the speed of sound in the conformal theory $v_s=1/\sqrt{3}$. In the left plot the values of $\kappa_1$ span from $-12.84$ (bottom curve) to $-5.18\times 10^3$ (upper curve). In the right plot we have marked the points where $\mu_r=\Lambda_\kappa$ for different curves, which are at the same values of $\kappa_1$ than in Fig.~\ref{fig:lambdak}. \label{fig:vstd} }	
	\end{center}
\end{figure}

\begin{figure}[h]
\begin{center}
\includegraphics[scale=0.45]{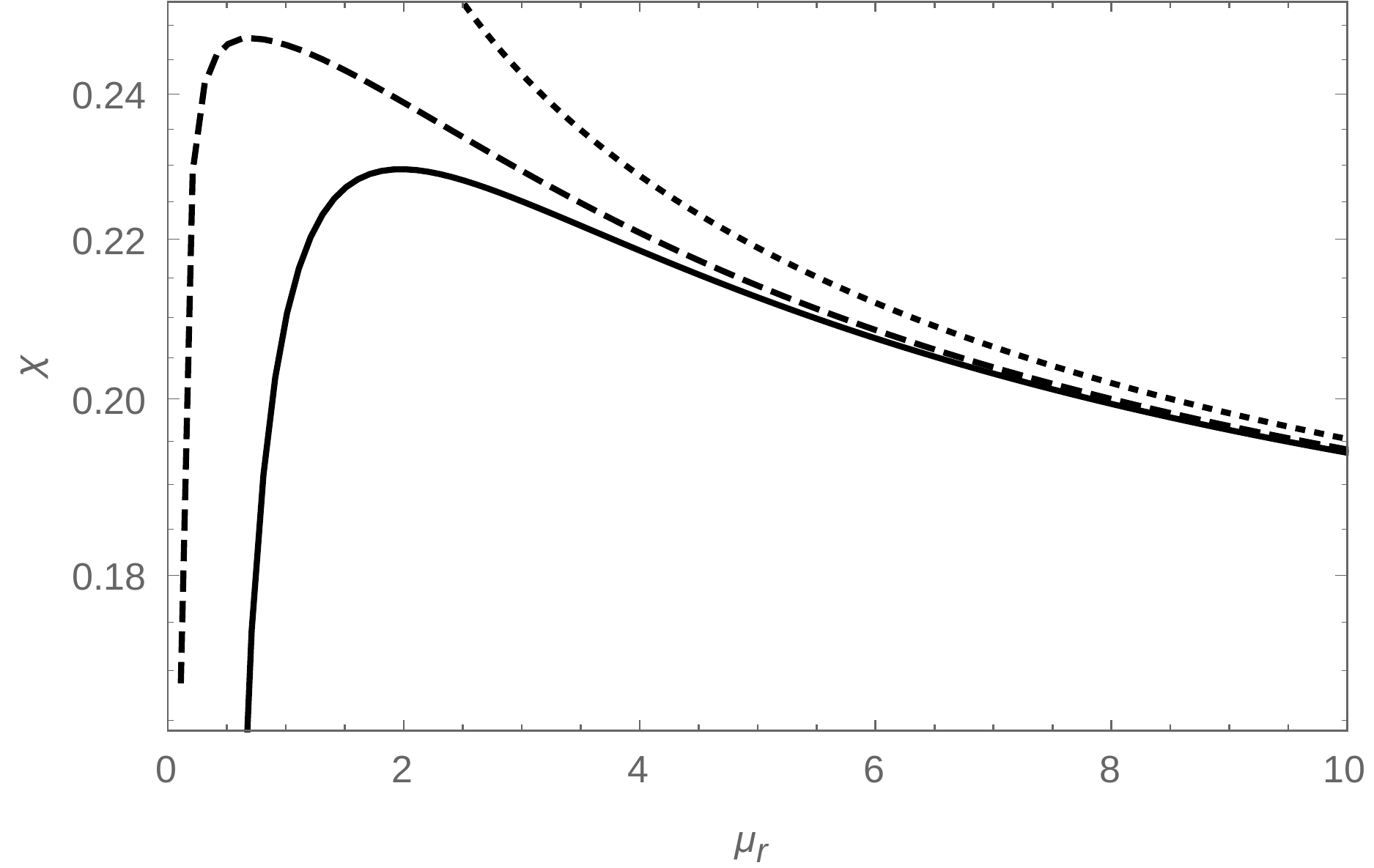} 
\caption{\small The charge susceptibility $\chi$ as a function of the reduced chemical potential $\mu_r$ for $t_r=1$ and for different values of $\kappa_1$. From top to bottom, $\kappa_1=-8$ (black dotted line), $\kappa_1 = \kappa_{\rm c}$ (black dashed line) and $\kappa_1 = -6$ (black solid line). The central curve marks the onset of the thermodynamic instability, i.e., $\chi(0)\vert_{\kappa=\kappa_{\rm c}}=0$. For larger values of $\kappa_1$, one gets $\chi<0$ up to some finite $\mu_r$. \label{fig:suscept}} 
\end{center}
\end{figure}

We can compare the values of $\tau$ in the models we study here with those of the degenerate Fermi liquid and the D3/D7 model, see Fig.~\ref{fig:Taub}. We observe that $\tau <0$ for a range of values of the chemical potential ($\mu_r\gtrsim 32$ for $\kappa_1=-10$; for even more negative values of $\kappa_1$ the curve of the top-down model goes further down). It thus appears that in these regions the EoS is stiffer than in a conformal theory, but how stiff can it be? In order to answer this question we would need to compute the adiabatic speed of sound \eqref{eq:vs}. However, it is technically easier to work at fixed temperature and compute the isothermal speed of sound
\be
v_{s\,\text{isot}}^2=\left(\frac{\partial p}{\partial \varepsilon}\right)_T= \frac{\left(\frac{\partial p_r}{\partial \mu_r}\right)_{t_r}}{\left(\frac{\partial \varepsilon_r}{\partial \mu_r}\right)_{t_r}} \ ,
\ee
which is closely related to the adiabatic one through the standard thermodynamic relations
\be
\begin{split}
&v_{s\,\text{isot}}^2=\frac{\rho_r}{\mu_r\left(\frac{\partial \rho_r}{\partial \mu_r}\right)_{t_r}+t_r\left(\frac{\partial s_r}{\partial \mu_r}\right)_{t_r}}\\
&v_{s\,\text{adiab}}^2=\frac{1}{\mu_r}\frac{\rho_r\left(\frac{\partial s_r}{\partial t_r}\right)_{\mu_r}-s_r\left(\frac{\partial s_r}{\partial \mu_r}\right)_{t_r}}{\left(\frac{\partial \rho_r}{\partial \mu_r}\right)_{t_r}\left(\frac{\partial s_r}{\partial t_r}\right)_{\mu_r}-\left(\frac{\partial \rho_r}{\partial t_r}\right)_{\mu_r}\left(\frac{\partial s_r}{\partial \mu_r}\right)_{t_r}}\ .
\end{split}
\ee
If the pressure has an analytic expansion in $T/\mu$ for $T/\mu \ll 1$ (as it will be the case in our models) and the entropy goes to zero at zero temperature, one can neglect the terms proportional to $\left(\frac{\partial s_r}{\partial \mu_r}\right)_{t_r}$ and the two speeds become the same. At non-zero temperature, the difference is suppressed by a factor of at least $O(T/\mu)$. Moreover, in many practical applications the temperature is taken to be zero as a good approximation. Therefore, we will study the isothermal speed of sound in the following and drop the label.

The behavior of the speed of sound $v_s$ in the top-down model is depicted in Fig.~\ref{fig:vstd}.\footnote{These plots correspond to values of the chemical potential that are much below the regime of validity of the probe approximation used in \cite{Hoyos:2016cob}. For $t_r=1$ one should go at least to values $\mu_r>150$ before we reach the probe limit. } We observe that, for $\kappa_1<0$, when $|\kappa_1|$ is increased the speed of sound becomes larger at low values of the chemical potential, eventually becoming quite close to the speed of light, and the region where the speed of sound is large also grows. A possible way to understand this is to recall that the scale $\Lambda_\kappa$ defined in \eqref{eq:lambdak} that controls the contribution of the logarithmic terms in \eqref{eq:renvevs} increases with increasing $|\kappa_1|$. When this happens, the logarithmic terms become large in magnitude. If the logarithmic terms in \eqref{eq:renvevs} dominate, the EoS becomes stiff but remains compatible with causality, as $\varepsilon_r\sim p_r$. Therefore, there is no fundamental obstacle towards obtaining a stiff EoS for a large interval of chemical potentials, as long as a significant separation of scales is present.

An important issue to consider is the possibility that the theory might become unstable in the stiff regime. A necessary but not sufficient condition for thermodynamic stability is that the charge susceptibility be positive, 
\be
\chi=\frac{\partial^2 p}{\partial \mu^2 } >0 \ .
\ee
In Fig.~\ref{fig:suscept} we plot $\chi\vert_{t_r =1}$ for different values of $\kappa_1$. For large enough values of $|\kappa_1|$, the susceptibility is positive and the theory is thermodynamically stable with respect to density fluctuations. There is a critical value $\kappa_1 = \kappa_{\rm c} \approx -6.44$, for which the theory becomes unstable at low values of the chemical potential. Therefore, the models with a large speed of sound are thermodynamically stable in the stiff regime. We will study their dynamical stability in Sec.~\ref{sec:stability}.

\subsection{Bottom-up models}

\begin{figure}[t]
	\begin{center}
	\begin{tabular}{cc}
		\includegraphics[scale=0.25]{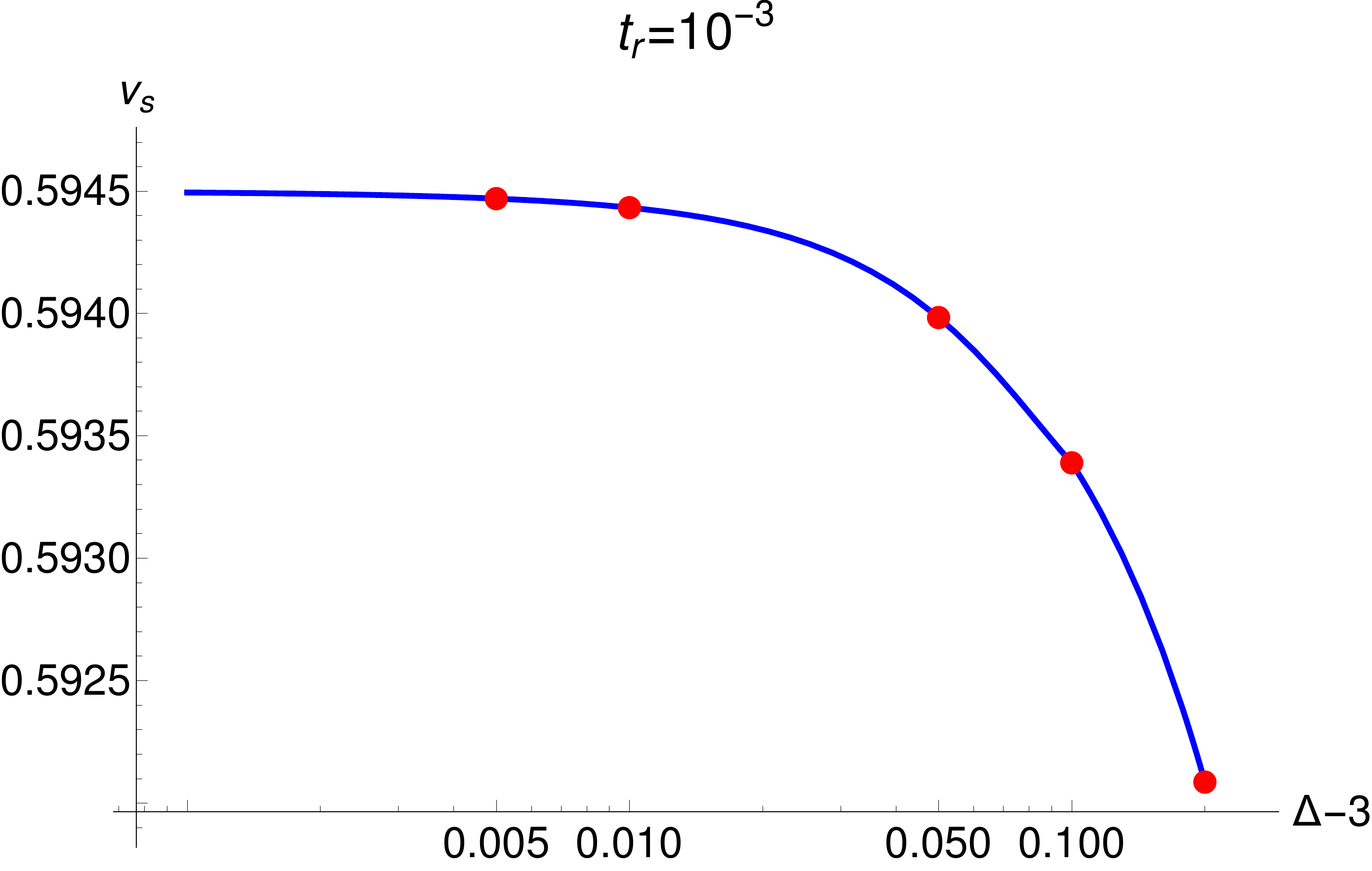} & \includegraphics[scale=0.25]{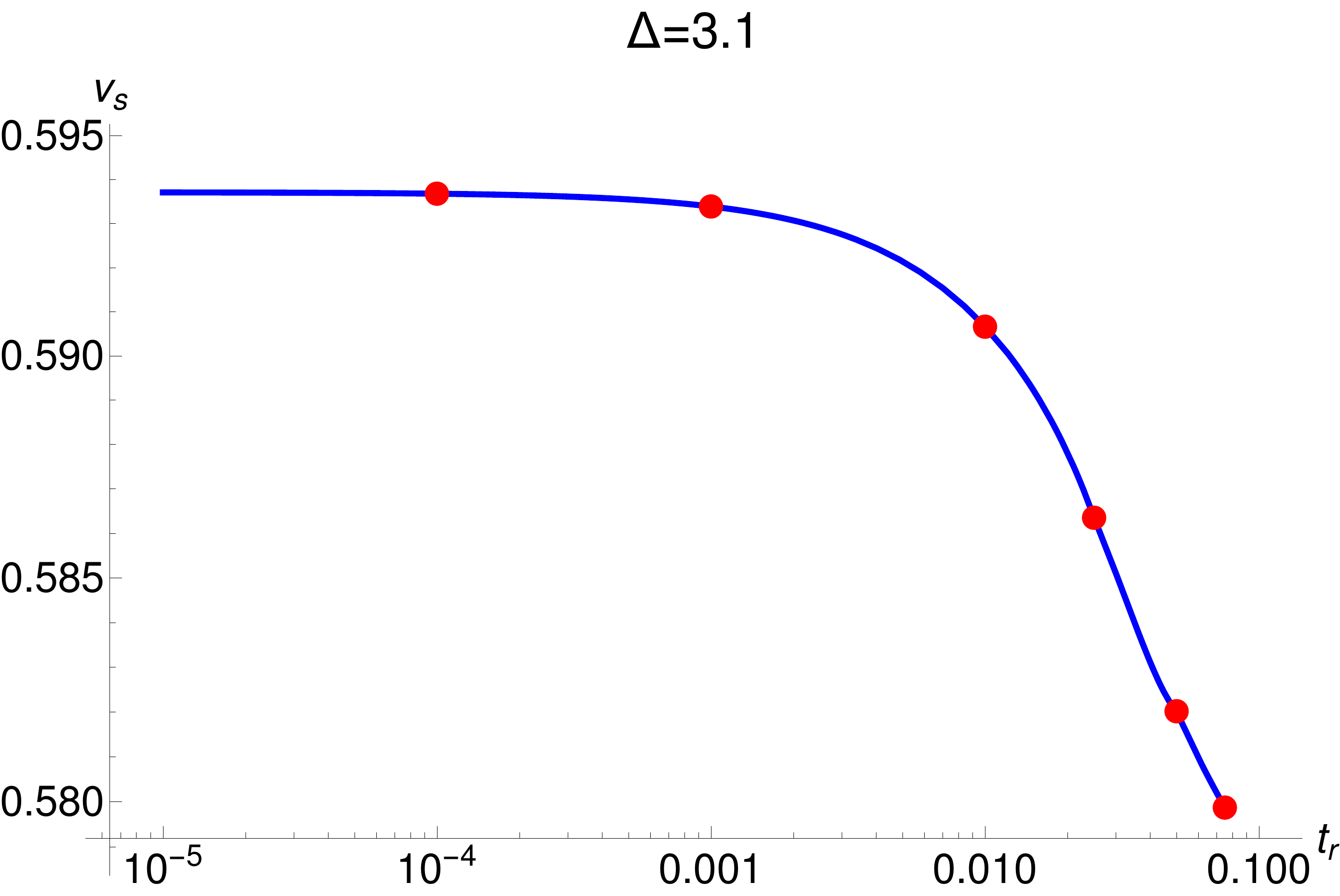}
		\end{tabular}
			\caption{\small Left plot: Maximum speed of sound at a given isotherm as a function of the conformal dimension. The isotherm was taken to be $t_r=10^{-3}$. Right plot: Maximum speed of sound as a function of the reduced chemical potential at a fixed conformal dimension $\Delta=3.1$.}\label{fig:maxvsV4} %In both cases, the saturation occurs when the deviation with respect to the conformal value is not greater than $0.1$.}
		\end{center}
\end{figure}

Moving again to the bottom-up models, we first consider the case without a quartic term in the potential $V_4=0$. The results are summarized in Fig.~\ref{fig:maxvsV4}. We find that the speed of sound can be larger than the one in a conformal theory, and that larger deviations occur for operators of lower dimensions, close to $\Delta=3$ for our allowed range. The left plot of Fig.~\ref{fig:maxvsV4} reflects this: there, we have fixed the temperature, computed the speed of sound as function of the chemical potential, and plotted the largest value we have found for each dimension of the scalar operator. This behavior holds for a range of low temperatures. The right plot of Fig.~\ref{fig:maxvsV4} shows the largest value of the speed of sound for a fixed dimension $\Delta\sim 3$ as we vary the temperature. We see that the magnitude increases as we lower the temperature, but it seems to saturate at an absolute maximum. The maximum value is just slightly larger than the conformal value by some $3\,\%$, while for phenomenological purposes it should be at least ca.~$30\,\%$ larger.

\begin{figure}[t]
\begin{center}
\begin{tabular}{cc}
 \includegraphics[scale=0.30]{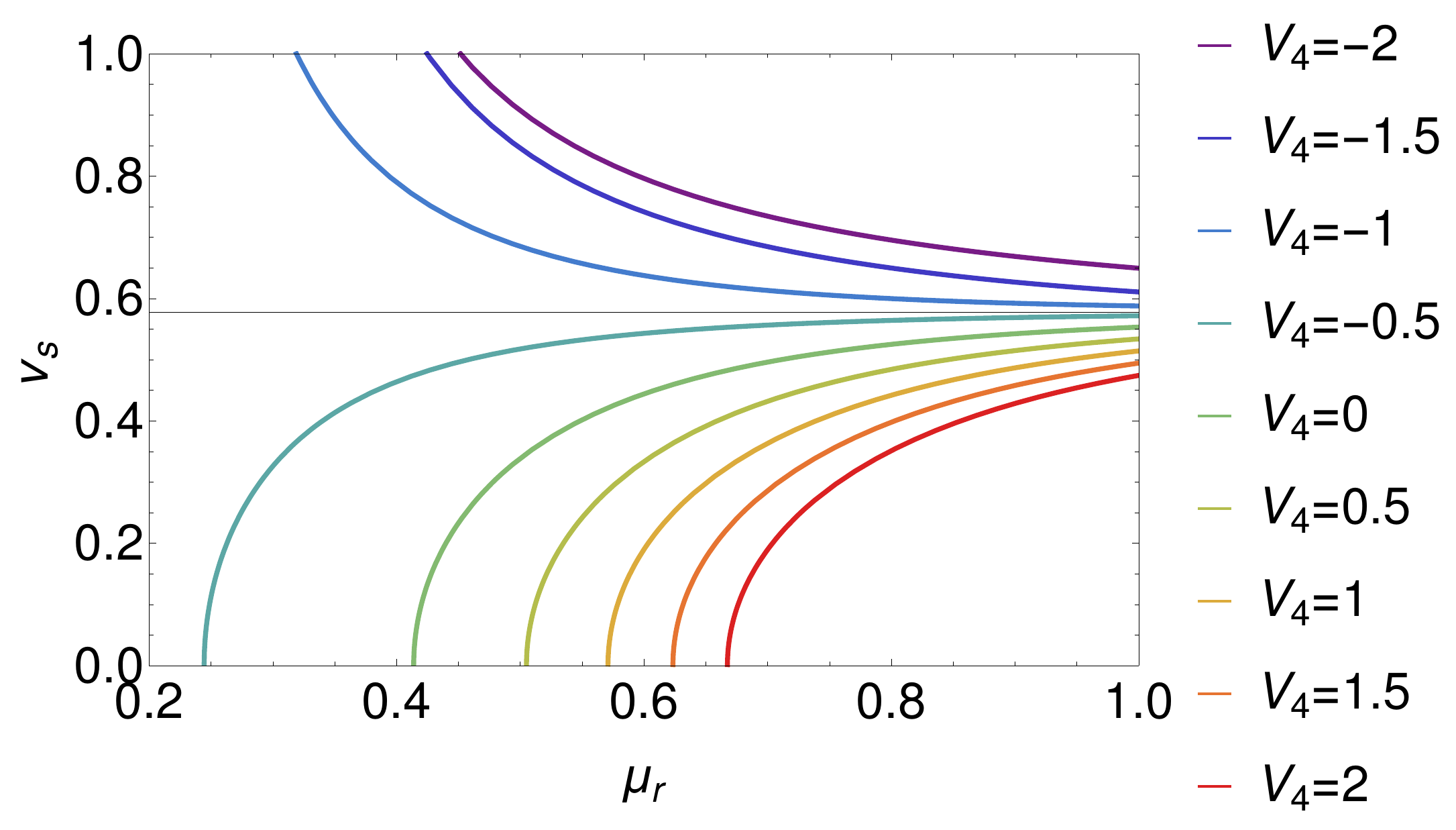} & \includegraphics[scale=0.30]{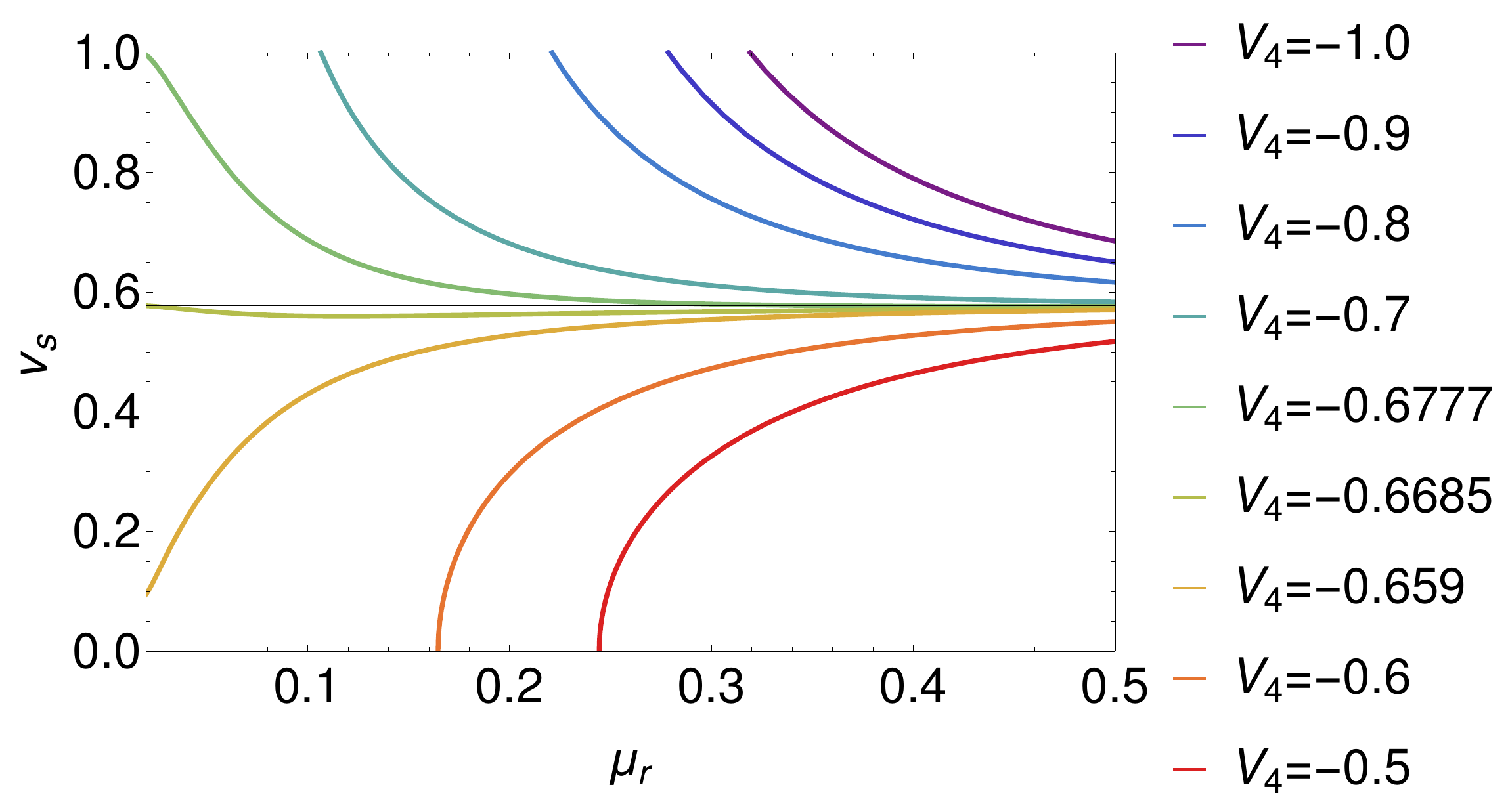} 
\end{tabular}
\caption{\small The speed of sound as a function of the reduced chemical potential $\mu_r$ for fixed temperature $t_r=0.1$. The charge and dimension of the dual scalar operator are $q=0$ and $\Delta=3$.
			}\label{fig:maxvs} %In both cases, the saturation occurs when the deviation with respect to the conformal value is not greater than $0.1$.}
\end{center}
\end{figure}

Next, we turn on the quartic term in the potential, i.e.~let $V_4\neq 0$. In Fig.~\ref{fig:maxvs} we plot the speed of sound as a function of the chemical potential for a fixed temperature $t_r=0.1$ and different values of $V_4$. We observe that making $V_4$ more negative increases the value of the speed of sound, while making $V_4$ more positive has the opposite effect. It is possible to reach values of the speed of sound $20-40\,\%$ larger than the conformal value for $V_4\sim -1.5$ and $\mu_r\sim 0.6-0.75$. The speed of sound seems to be growing further at lower values of the chemical potential. This shows that stiff phases are possible in generic holographic models. 

\begin{figure}[t]
\begin{center}
\begin{tabular}{cc}
 \includegraphics[scale=0.30]{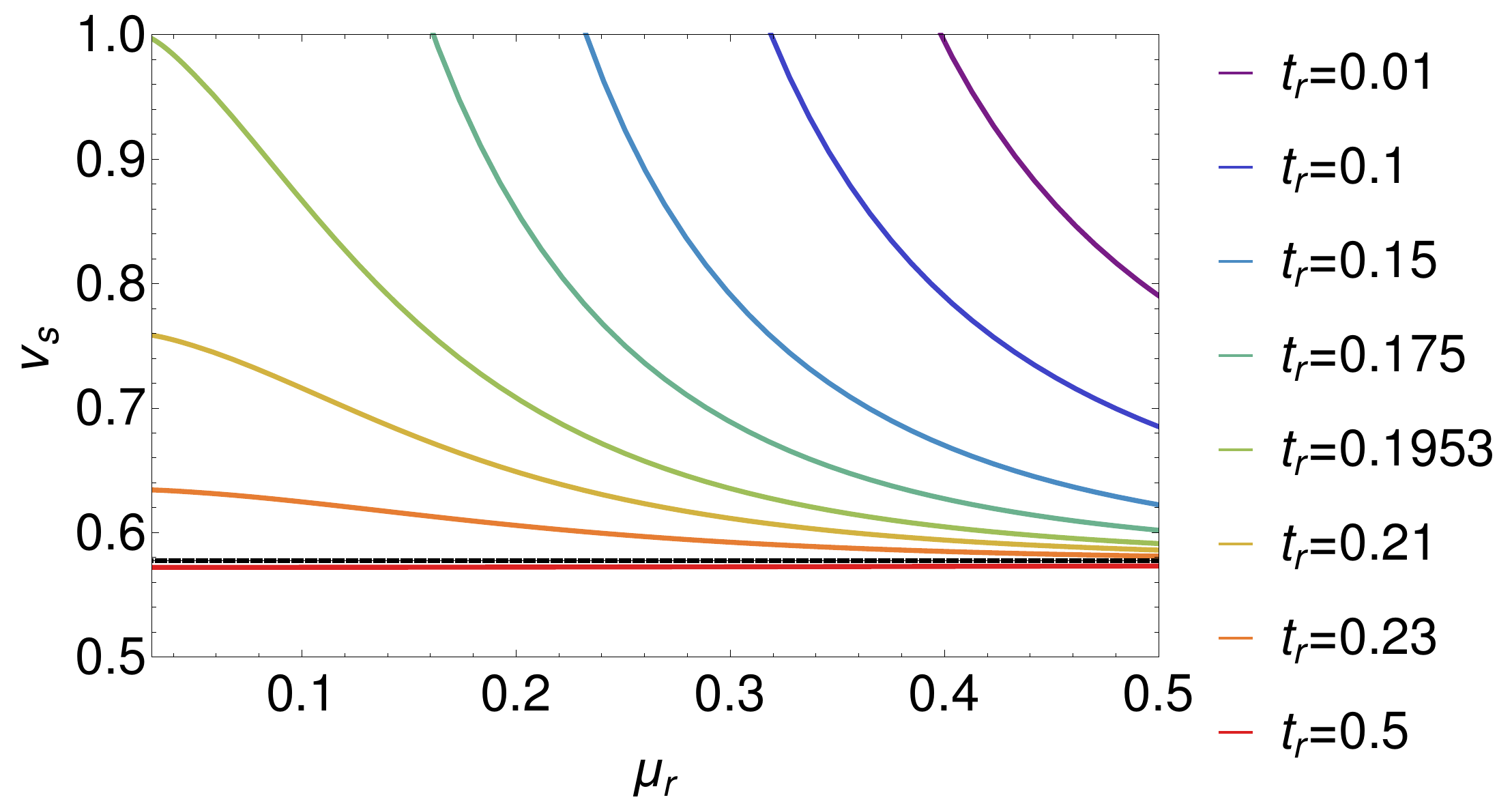} & \includegraphics[scale=0.30]{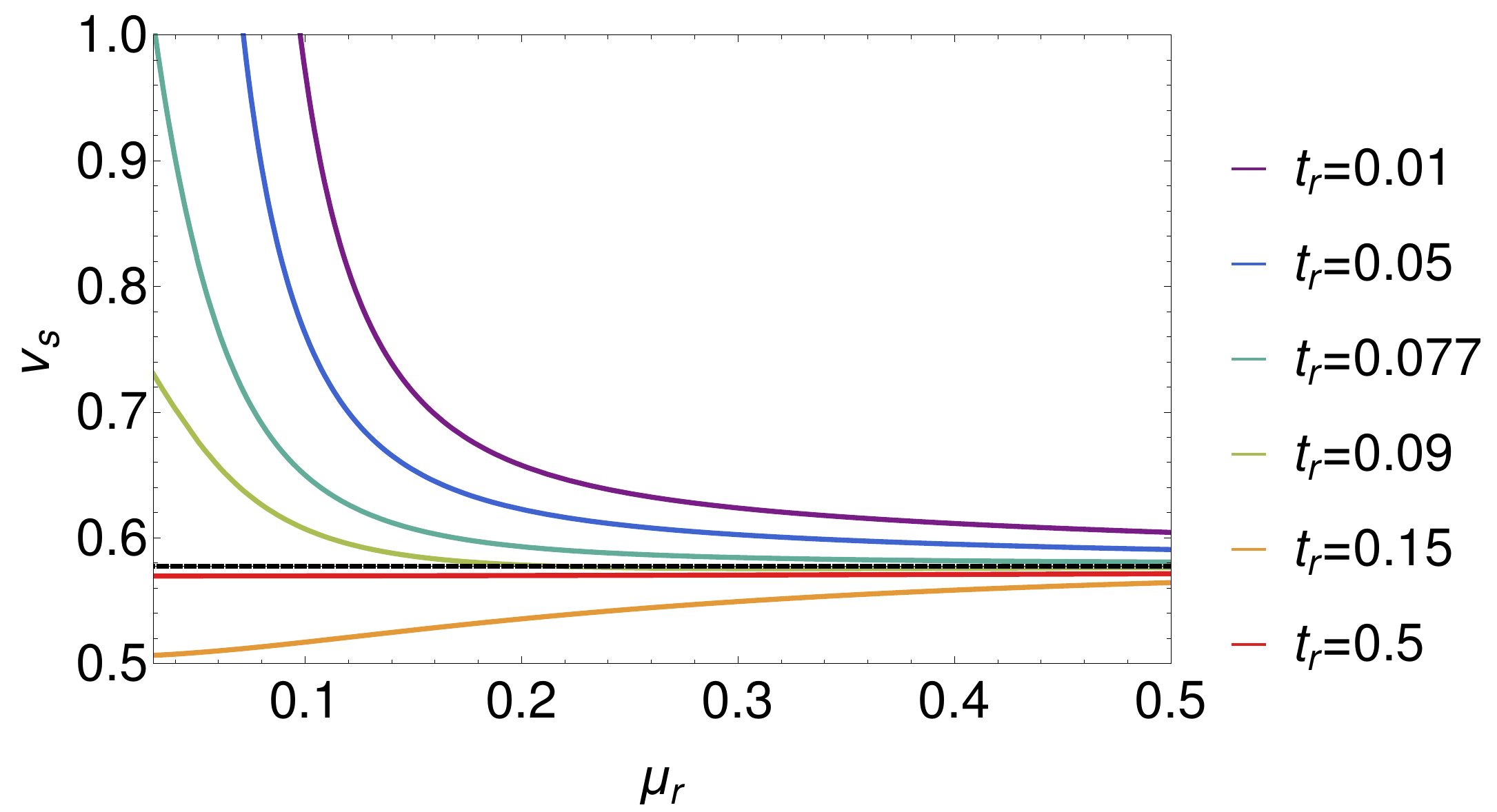} 
\end{tabular}
\caption{\small The speed of sound as a function of the reduced chemical potential $\mu_r$ for different temperatures and for a fixed quartic potential $V_4=-1$ (left) and $V_4=-0.67$ (right). The charge and dimension of the dual scalar operator are $q=0$ and $\Delta=3$.
\label{fig:vsOverTbottomUp}} 
\end{center}
\end{figure}

However, in contrast to the top-down model, we find that in most cases there are violations of causality ($v_s>1$) or thermodynamic instabilities ($v_s^2<0$) at small values of the chemical potential, so there is likely a phase transition between the high temperature, zero density phase and the low temperature, non-zero density one. Nevertheless, as we show in Fig.~\ref{fig:vsOverTbottomUp}, for any given temperature there is a range of values of $V_4$ where the speed of sound remains in the physical range $1\geq v_s^2\geq 0$. This happens around the special value of $V_4=-2/3$, for which the conformal anomaly vanishes; we even observe that near the special value the speed of sound becomes very close to its conformal limit and almost independent of the chemical potential. 
%We believe that more precise numerics would show that the value of $V_4$ for which the speed of sound becomes constant is indeed $V_4=-2/3$.

%\begin{figure}[htb!]
% \begin{center}
%  \includegraphics[width=12cm]{vs2BottomUp.pdf} 
%			\caption{\small Around to the special value $V_4=-2/3$ we observe that the speed of sound remains in the physical window $1\geq v_s^2\geq 0$. For $V_4\simeq-0.6685$  the speed of sound becomes independent of the chemical potential and its value is the same as in a conformal theory. The temperature for all the curves is fixed to $t_r=0.1$.% and $\kappa_2=-1.80,-0.67,-0.34,-0.42,-0.66,-0.99,-1.36,-1.78,-2.22$.
%			}\label{fig:V4wind} %In both cases, the saturation occurs when the deviation with respect to the conformal value is not greater than $0.1$.}
%		\end{center}
%\end{figure}

\begin{figure}[t]
\begin{center}
\begin{tabular}{cc}
 \includegraphics[scale=0.30]{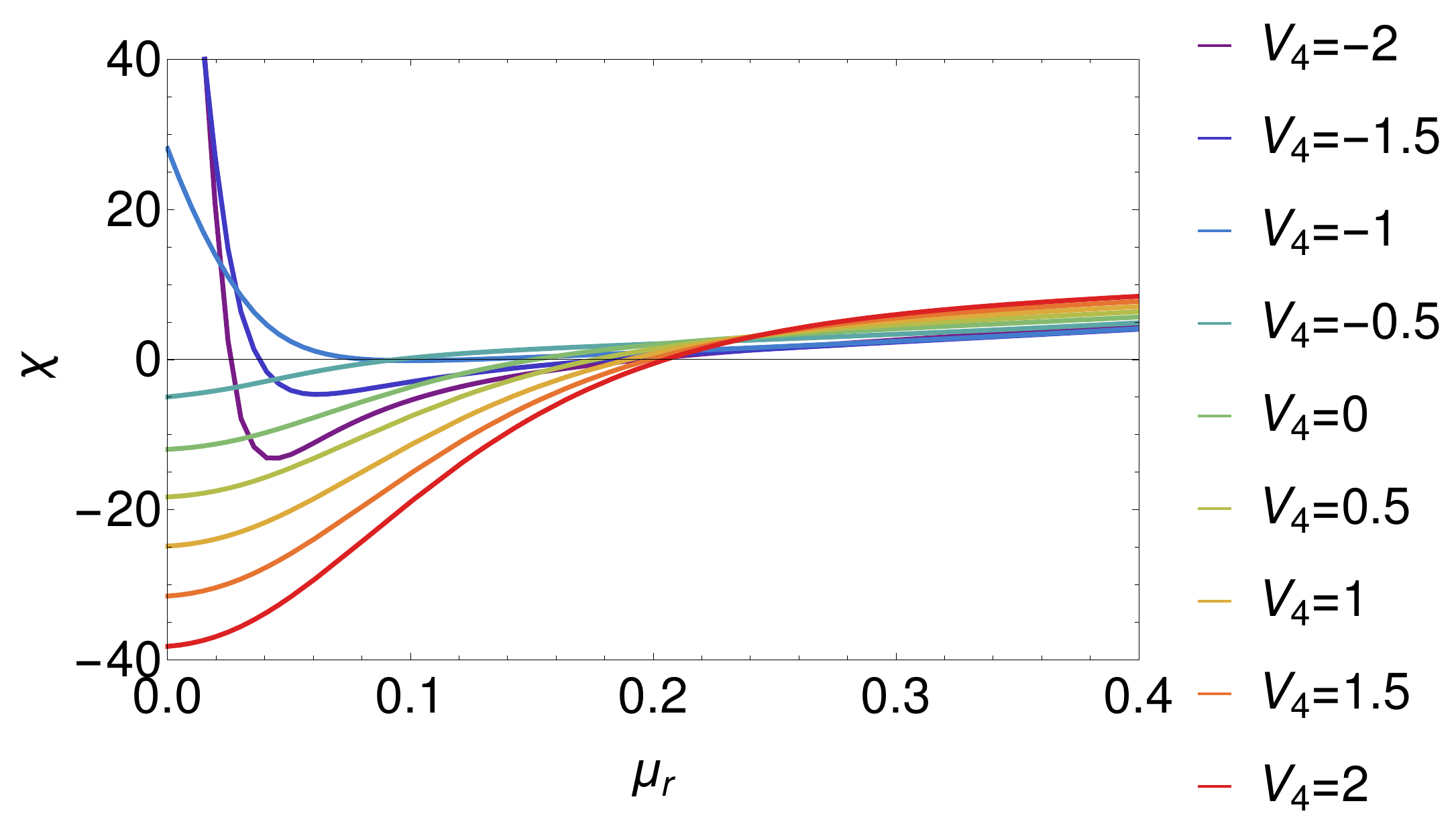} & \includegraphics[scale=0.30]{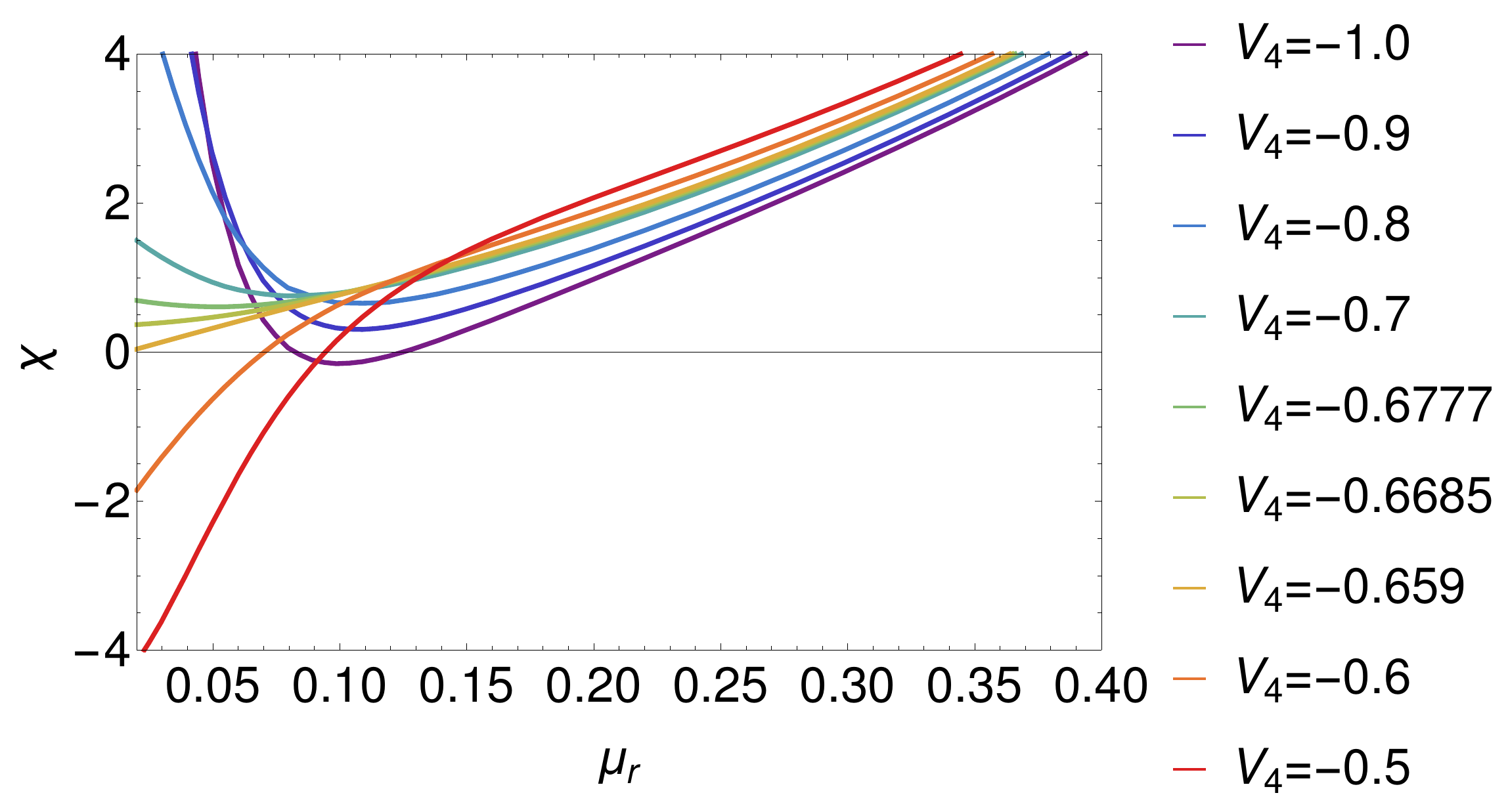} 
\end{tabular}
\caption{\small The charge susceptibility $\chi$ as a function of the reduced chemical potential $\mu_r$ for $t_r=0.1$ and for different values of $V_4$. \label{fig:susceptbot}} 
\end{center}
\end{figure}

To inspect thermodynamic stability, we have plotted the charge susceptibility in Fig.~\ref{fig:susceptbot}. In the cases where $v_s^2<0$ we also find that the susceptibility becomes negative, although this appears to happen at lower values of the chemical potential, so it may correspond to a different kind of instability. When the speed of sound becomes superluminal there can also be a small interval with $\chi>0$ for $V_4\gtrsim -1$. In the window where $1\geq v_s^2> 0$ for all values of the chemical potential we find that $\chi>0$, so these correspond to thermodynamically stable phases. Numerically, it seems that the values of $V_4$ for which $v_s^2=0$ and $\chi=0$ at zero chemical potential coincide.

\section{Stability}\label{sec:stability}

The aim of this section is to determine whether the stiff phases we have found are indeed local minima of the free energy in the space of homogeneous configurations. To this end, we will introduce a small time-dependent perturbation, expecting that if the equilibrium configuration is unstable we will witness the exponential growth of some of the modes. Otherwise, we expect the perturbation to oscillate and/or decay back to equilibrium. As we are considering only homogeneous configurations, we can suppress the spatial dependence. On the gravity side, this translates into studying a linear perturbation around the previously obtained background solution\footnote{We work in the $\delta g_{r\mu}=A_r=0$ gauge.}
\be 
\Phi \to \Phi(r) + \delta \Phi(r,t), \quad A_0 \to A_0(r) + \delta A_0(r,t) , \quad g_{\mu\nu} \to g_{\mu\nu}(r) + \delta g_{\mu\nu}(r,t) \ .
\ee
Since our background is stationary, we can expand in plane waves of a given frequency $\omega$,
\be 
\delta \Phi(r,t) = \varphi(r) e^{-i\omega t} , \quad \delta A_0(r,t) = a_0(r) e^{-i\omega t}, \quad \delta g_{\mu\nu}(r,t) = h_{\mu\nu} (r)e^{-i \omega t}\ .
\ee
Dynamical modes are normalizable and satisfy an ingoing boundary condition at the horizon. This is possible typically only for a discrete set of complex frequencies, the quasinormal frequencies $\omega_n$. If the imaginary part of the quasinormal frequency is negative or zero, ${\rm Im}\,\omega_n\leq 0$, the associated quasinormal mode decays in time or is oscillatory, and the background is stable. On the other hand, if ${\rm Im}\,\omega_n> 0$ is positive, the quasinormal mode grows exponentially in time and the background is unstable.

An important piece of information is that at zero chemical potential and high temperatures --- $\mu_r=0$, $t_r\gg 1$ --- the model is known to be stable, as the quasinormal modes should approximate those of a probe scalar in an $AdS$ black hole background, all of which are on the lower half of the complex frequency plane \cite{Nunez:2003eq}. As the background changes continuously, a quasinormal mode has to cross the real axis to the upper half plane in order to develop an instability. Physically we expect that if the background becomes unstable there will be another {\em stationary} solution corresponding to the true vacuum of the theory. In that case the crossing should happen at the origin of the complex plane. Therefore, the onset of the instability can be determined from the appearance of a quasinormal mode at zero frequency.

\begin{figure}[t]
\begin{center}
\begin{tabular}{cc}
 \includegraphics[scale=0.35]{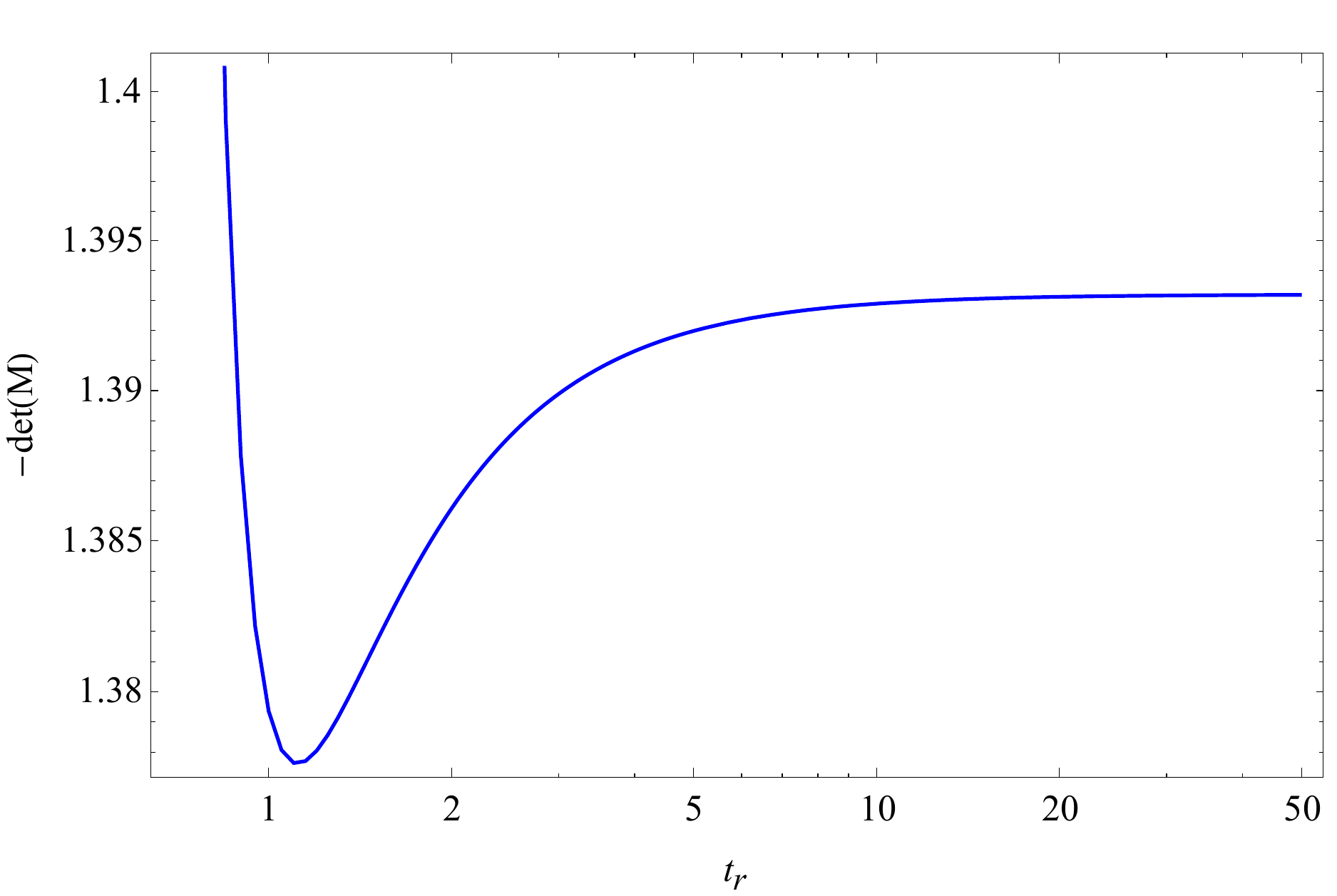} & \includegraphics[scale=0.35]{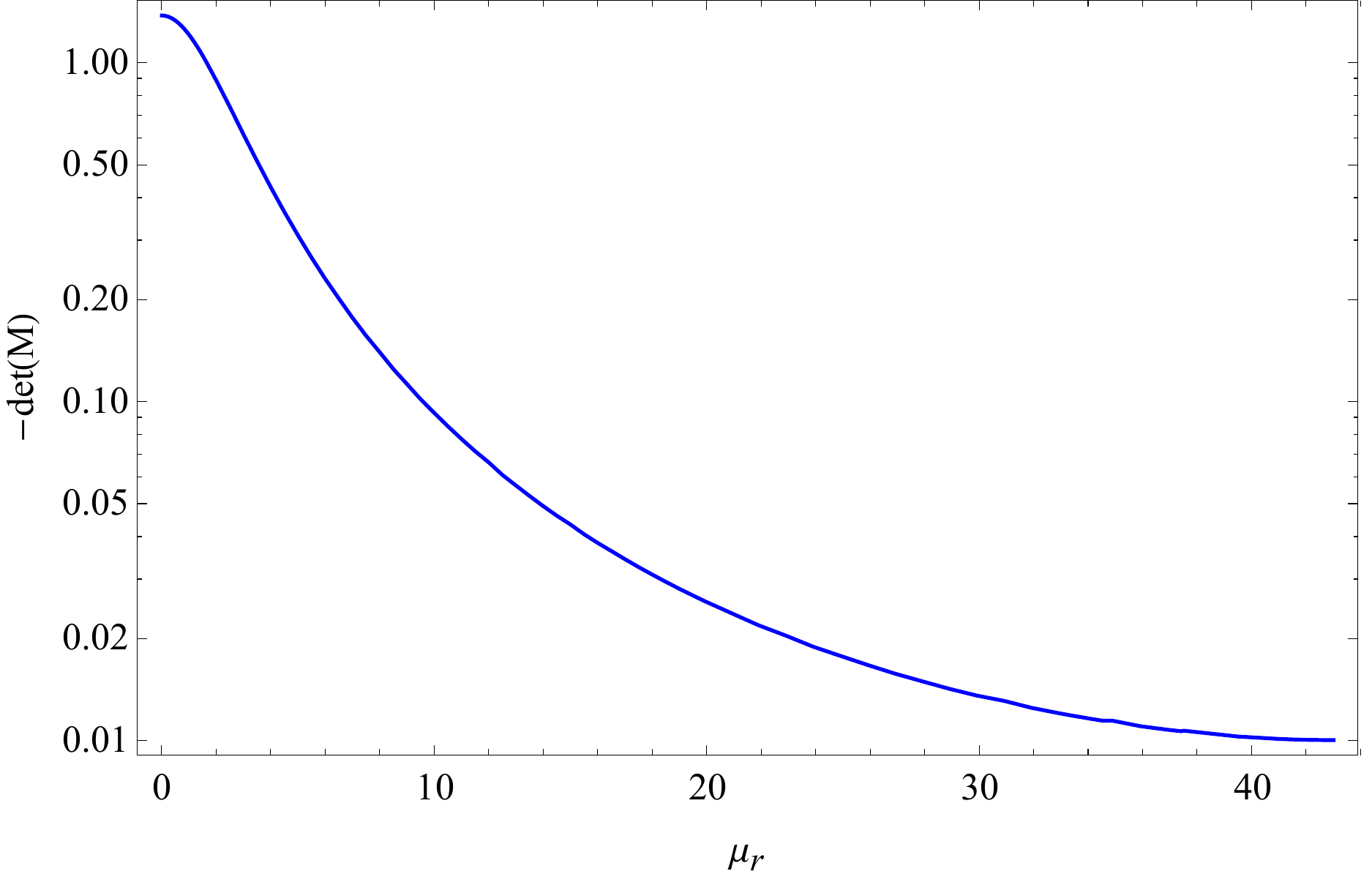} 
\end{tabular}
\caption{\small Left figure: $\text{det}(M)$ at zero density as a function of the reduced temperature. Right figure: $\text{det}(M)$ as a function of the reduced chemical potential at $t_r=1$.  \label{fig:stablezerod}}
\end{center}
\end{figure}

We study the appearance of a zero frequency quasinormal mode using the determinant method of e.g.~\cite{Amado:2009ts,Bergman:2011rf} and standard techniques, most details of which can be found in Appendix~\ref{app:numerics}. First, we introduce gauge invariant combinations of the fields under diffeomorphisms that preserve the condition $g_{\mu r}=0$. There are two independent scalar modes
\be\label{eq:zs}
\begin{split}
z_1=& \varphi+\varphi^\dagger-\frac{r \Phi_0'}{1+r A'}h \\
z_2=& \omega \left(\varphi-\varphi^\dagger\right)+q\Phi_0\left[\frac{A_0}{f}h_{00}+2 a_0+\left(A_0+\frac{r}{1+r A'}\left(A_0'-\frac{f'}{2f}A_0\right)\right)h\right]\ ,
\end{split}
\ee
where $h=\delta^{ij}h_{ij}/3$ is the trace of the spatial components of the metric fluctuation. If $q\neq 0$, the two modes are coupled 
\be
\begin{split} \label{eq:eomsfluc}
0 &= z_i''+  \mathcal{A}_{ij} z_{j}  + \mathcal{B}_{ij} z_j' , \qquad i,j = 1,2\ , \\
\end{split}
\ee
with coefficients $\mathcal{A}_{ij},\mathcal{B}_{ij}$ that depend on the background fields. If $q=0$, the off-diagonal components of $\cA$ and $\cB$ are zero and the two modes decouple. 

We impose that the solutions are ingoing at the horizon. There are two independent solutions $z_i^{(I)}$, $z_i^{(II)}$ corresponding to making $z_1$ or $z_2$ zero at the horizon. When these solutions are taken to the boundary, a linear combination of them will be normalizable for the values of the frequency corresponding to the quasinormal modes. In the $u$ coordinate, the expansion of the solutions at the boundary $u\to 0$ is, to leading order,
\be 
z_i \sim \sqrt{u}\left(z_i^{(nn)} + u\, z_i^{(n)} \right) \ ,
\ee 
where we identify the coefficients of the non-normalizable $(nn)$ and normalizable $(n)$ solutions. We arrange the solutions in a matrix with constant entries at the boundary
\be 
M=\lim_{u\to 0} \frac{1}{\sqrt{u}}\left(
\begin{array}{c c}
z_1^{(I)} & z_2^{(I)}\\
z_1^{(II)}  & z_2^{(II)}\\
\end{array}
\right)\ .
\ee
$M$ depends on the frequency, and a normalizable solution exists when $M(\omega)$ has a zero eigenvalue, i.e.~$\det( M(\omega))=0$. 

For the top-down model we have computed the determinant at zero frequency $\omega=0$ first for a zero chemical potential $\mu_r=0$ starting at high temperatures and decreasing the temperature to values $t_r<1$ (left plot in Fig.~\ref{fig:stablezerod}). As the determinant never vanishes, the background is stable for $\mu_r=0, t_r=1$. We then repeat the same calculation but keeping $t_r=1$ fixed and increasing the chemical potential $\mu_r$. We find that the determinant is non-vanishing in the range we are interested $0\leq \mu_r\leq 50$ (right plot in Fig.~\ref{fig:stablezerod}). Therefore, the theory remains dynamically stable in the regime where the EoS is stiff.

\begin{figure}[t]
\begin{center}
\begin{tabular}{cc}
  \includegraphics[scale=0.27]{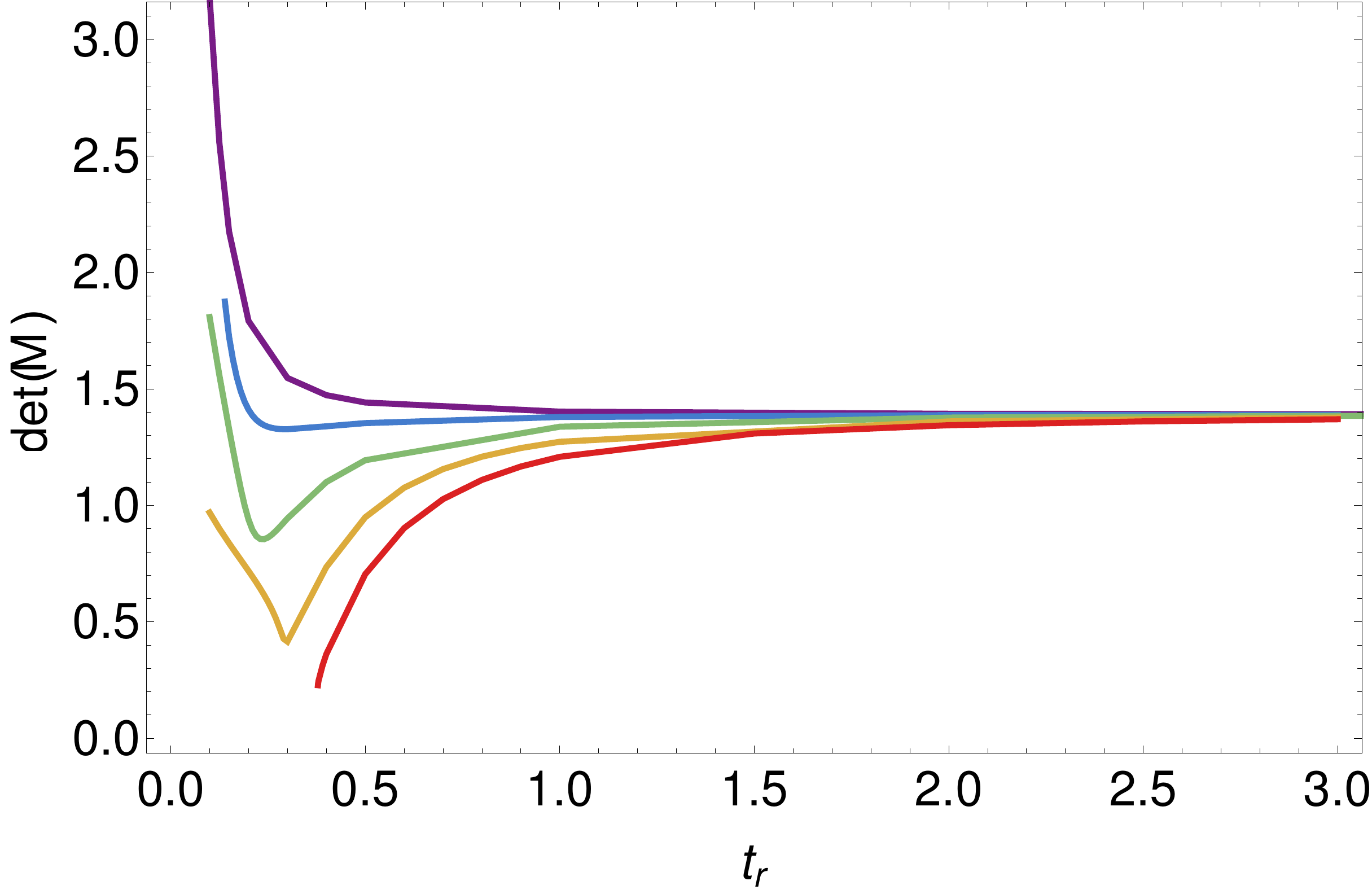} &\quad \includegraphics[scale=0.27]{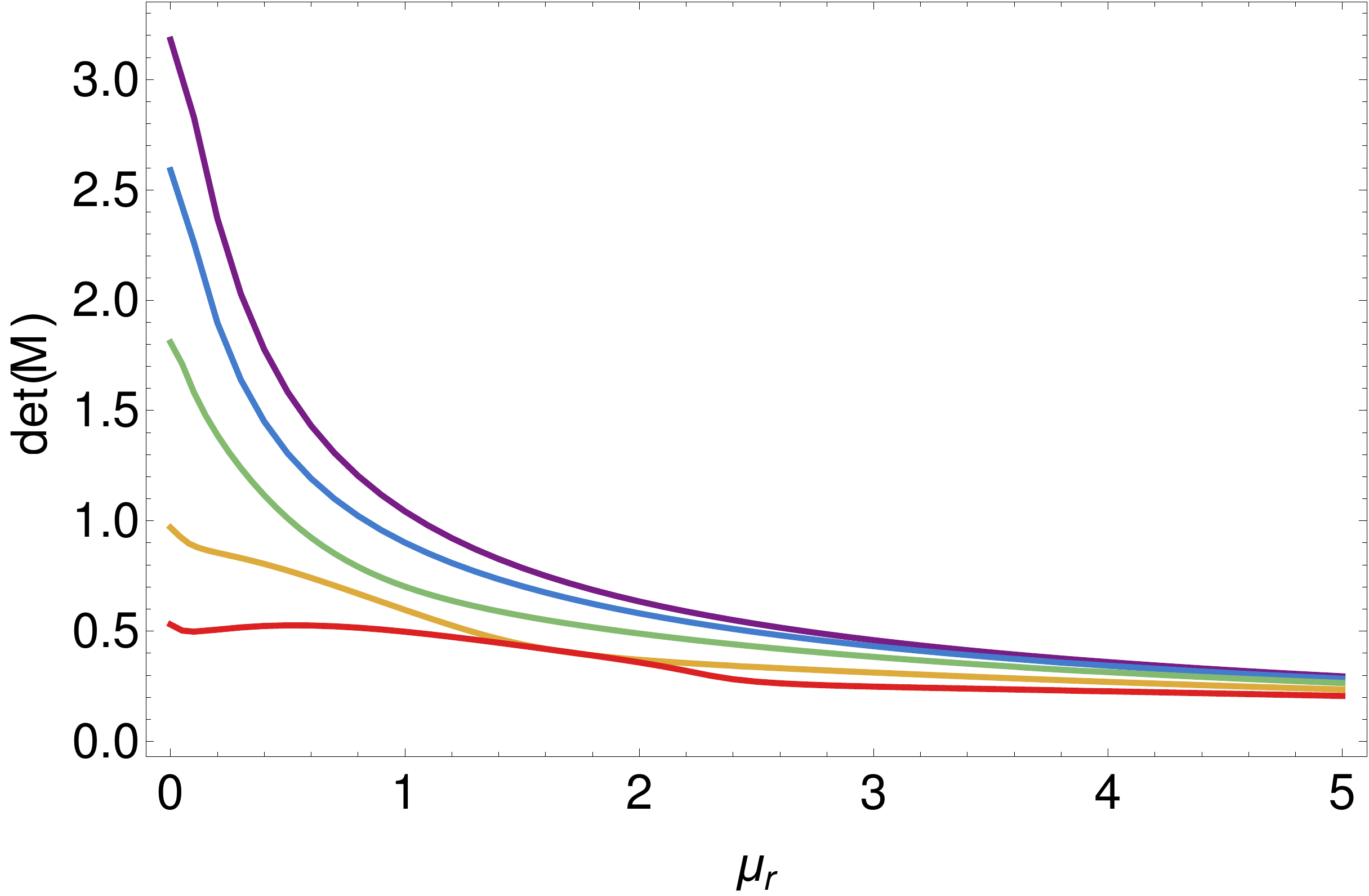}
\end{tabular}
\caption{\small Left figure: $\text{det}(M)$ at zero density as a function of the reduced temperature for the bottom-up model with quartic term (top to bottom) $V_4=-0.5,-0.6777,-1,-1.5,-2$. Right figure: $\text{det}(M)$ as a function of the reduced chemical potential at $t_r=0.1$.  \label{fig:stablezeroBU}}
\end{center}
\end{figure} 

For the bottom-up model we do a similar stability analysis, we first compute the determinant at zero frequency $\omega=0$ at zero chemical potential $\mu_r=0$ starting at high temperatures and decreasing the temperature to values $t_r<0.1$ (left plot in Fig.~\ref{fig:stablezeroBU}). We then fix the temperature to $t_r=0.1$ and increase the chemical potential $\mu_r$ (right plot in Fig.~\ref{fig:stablezeroBU}). We find that the determinant is non-vanishing for values of $V_4$ where the speed of sound remains in the physical window, even when the speed of sound is close to the speed of light. Therefore,  these models have sensible physical behavior and no obvious instabilities even in the regime where the EoS is stiff.

\section{Conclusions}\label{sec:conclusions}

In the paper at hand, we studied the thermodynamics of cold and dense strongly coupled matter via simple holographic models. The models include the minimal ingredients of finite charge density and breaking of conformal invariance through a coupling for a relevant scalar operator of conformal dimension $4>\Delta\geq 3$. We find that for some of these cases it is possible to find very stiff Equations of State, with the speed of sound almost reaching the speed of light. A simple stability analysis of the models furthermore showed no obvious thermodynamic or dynamic instabilities.

We observe that the simplest models possessing a quadratic action for the scalar field do not reach speeds of sound significantly larger than the conformal limit of $v_s=1/\sqrt{3}$. In bottom-up models with a quartic potential, the speed of sound can on the other hand reach the speed of light if the quartic term has a negative coefficient $V_4<0$ with large enough magnitude. However, except for a small range of values around 
$V_4=-2/3$, the isothermal speed of sound becomes superluminal or imaginary (indicating the presence of an instability) at low values of the chemical potential. Concerning the superluminal behavior, it should, however, be noted that when the chemical potential is of the same order or smaller than the temperature, one should rather consider the adiabatic speed of sound, which may affect to the range of values of $V_4$, for which causality is respected.

In addition to the bottom-up models, we also studied a top-down model with a more complicated action for the scalar, determined by a consistent truncation of supergravity. The issues of superluminal or imaginary speeds of sound do not appear in this case, which suggests that adding higher powers of the scalar field to the scalar potential might ameliorate the behavior of these quantities also in the bottom-up models. On the other hand, a stiff EoS is achieved in the top-down model only when there is a large separation between the scale of explicit breaking of conformal invariance and another scale that is spontaneously  generated due to logarithmic divergences. The conclusion seems to be that although there is no fundamental obstruction to achieving a stiff EoS, this may not be possible in the simplest models and/or for the most ``natural'' values of the parameters of the system. On the positive side, the conditions required to achieve a stiff and physically consistent EoS may prove to be quite restrictive and thus turn out to be useful in constraining possible holographic models of QCD.

An obvious phenomenological application of our results lies in the physics of neutron stars, where a holographic quark matter EoS has previously been matched to nuclear matter EoSs in \cite{Hoyos:2016zke}. The fact that very stiff EoSs can be obtained from holography opens up the possibility to construct matched EoSs exhibiting a weakly first order or even a cross-over deconfinement transition, thus allowing for the existence of a macroscopic amount of quark matter in the cores of the stars. Recalling the ease, with which quantities such as neutrino emissitivities and transport coefficients can be computed in holography, this paves the way for very interesting astrophysical studies.

\section*{Acknowledgments}
We would like to thank Daniel Grumiller, Alexander Haber, Ville Ker\"anen, Esko Keski-Vakkuri, Elias Kiritsis, Aleksi Kurkela, David M\"uller, Ayan Mukhopadhyay, Francesco Nitti, Florian Preis, Anton Rebhan, Luciano Rezzolla, and Andreas Schmitt for useful discussions.
C.~E.~is supported by the Austrian Science Fund (FWF), project no. P27182-N27 and DKW1252-N27.
C.~H.~is supported by the Ramon y Cajal fellowship RYC-2012-10370 and the Spanish national grant MINECO-16-FPA2015-63667-P. 
C.~H.~and D.~R.~F.~are supported by the Asturian grant FC-15-GRUPIN14-108.
N.~J.~is supported by the Academy of Finland, grant no.~1297472.
A.~V.~is supported by the Academy of Finland, grant no.~1303622, as well as the European Research Council, grant no.~725369.
N.~J. and A.~V. thank the organizers of the CERN workshop \emph{From quarks to gravitational waves: Neutron stars as a laboratory for fundamental physics} for warm hospitality.

\newpage 
\appendix

\section{Background solutions}\label{app:eoms}

Varying the action \eqref{eq:langdens} of the top-down model \eqref{eq:topdownL} with respect to the bulk metric, gauge, and scalar fields yields the equations of motion
\be 
\begin{split} \label{eq:eoms}
0 &=\frac{\Phi '}{f r} \left[ f \left(4 r A'+5\right)+r f'\right] -\frac{\Phi ^2+1  }{ \Phi
	^2-1 }\frac{e^{-2 A} \Phi}{f^2 r^4} \left(3 e^{2 A} f r^2+A_0^2 L^4 q^2\right) +\frac{2 \Phi  \left(\Phi '\right)^2}{1- \Phi ^2} +\Phi '' \\
0 &=A_0' \left(2 A'+\frac{3}{r}\right) -\frac{4 A_0 q^2 \Phi ^2}{f r^2 \left(1-\Phi
	^2\right)^2}+A_0'' \\
0 &=f' \left(4 A'+\frac{5}{r}\right) -\frac{16 e^{-2 A} A_0^2 L^4 q^2 \Phi ^2}{f r^4 \left(1-
	\Phi ^2\right)^2}-\frac{4 L^4}{r^2}  e^{-2 A} A_0'{}^2 +f'' \\
0 &=A''+ \frac{A'}{r} + \frac{8}{3 \left(1- \Phi ^2\right)^2 } \left(  \Phi '^2 +   \frac{L^4 q^2}{ f^2 r^4} e^{-2 A} A_0^2  \Phi ^2 \right) \\
0 &=  A' \left(\frac{3 f'}{2 f}+\frac{12}{r} + 6A' \right) +\frac{1}{2f r^2} \left[ 2 e^{-2 A} L^4 A_0'{}^2 + 3 \left(r f'-4 \frac{1+ \Phi^4}{\left(1-\Phi ^2\right)^2} \right) \right] +\\
 & \frac{6}{r^2}-\frac{4 }{\left(1-\Phi ^2\right)^2}\left(  \Phi '^2 +\frac{L^4 q^2 }{f^2 r^4} e^{-2 A} A_0^2  \Phi ^2\right)  \ . \\
\end{split}
\ee

\subsection{Near boundary series expansions}
\label{app:nbseries} 

The near boundary behavior for the scalar field is
\be 
\Phi \sim \frac{L^2}{r}\phi_{(0,1)} + \frac{L^6}{r^3}\left[ \phi_{(1,3)} \log\left(\frac{r}{L}\right) + \phi_{(0,3)}\right]\ .
\ee
We shall assume the following series expansions for the other fields 
\be 
\begin{split}
f & = 1+ \sum_{n,m} \frac{L^{2n}}{r^n}f_{(n,m)}\log \left(\frac{r}{L}\right)^m, \quad A = \sum_{n,m} \frac{L^{2n}}{r^n}A_{(n,m)}\log \left(\frac{r}{L}\right)^m \\
 A_0 &= \mu + \sum_{n,m} \frac{L^{2n}}{r^n}A_0{}_{(n,m)}\log \left(\frac{r}{L}\right)^m, \quad \Phi = \sum_{n,m} \frac{L^{2n}}{r^n}\phi_{(n,m)}\log \left(\frac{r}{L}\right)^m
\end{split} 
\ee
which upon implementing the equations of motion become 
\be 
\begin{split} \label{eq:nbseries}
A_0(r) &\sim \mu +\frac{L^4}{r^2}\left[\text{A}_{0}{}_{(0,2)}-8 \mu \phi_{(0,1)}^2 \log\left(\frac{r}{L}\right)\right] + \\
&  \frac{L^8}{3 r^4}\Bigg\lbrace 2\phi_{(0,1)} \left[\phi_{(0,1)} \left(4 \text{A}_0{}_{(0,2)}+9 \mu ^3\right)-8 \mu \phi_{(0,1)}^3+6 \mu  \phi _{\text{(0,3)}}\right] +\\
& 24 \log \left(\frac{r}{L}\right) \mu \phi_{(0,1)}^2 \left(\mu ^2\phi_{(0,1)}^2-2 \phi_{(0,1)}^2\right)\Bigg\rbrace \\
f(r) &\sim  1+ \frac{L^8}{r^4}\left[f_{0,4}-16 \mu ^2 L^4\phi_{(0,1)}^2 \log \left(\frac{r}{L}\right)\right] \\
A(r) &\sim -\frac{2 L^4\phi_{(0,1)}^2}{3 r^2} -\frac{L^8}{9r^4}\phi_{(0,1)} \Bigg\lbrace 9 \mu ^2\phi_{(0,1)} \left[2 \log\left(\frac{r}{L}\right)+1\right] +\phi_{(0,1)}^3\left[12 \log \left(\frac{r}{L}\right)+5\right]+9 \phi_{\text{(0,3)}} \Bigg\rbrace  \\
\Phi(r) &= \frac{L^2}{r}\phi_{(0,1)}  + \frac{L^6}{r^3}\left[\phi_{(0,1)} \log \left(\frac{r}{L}\right) \left(\frac{4}{3}\phi_{(0,1)}^2+ 2\mu ^2 \right)+ \phi_{\text{(0,3)}} \right]
\end{split}
\ee
plus sub-leading terms that we do not put here.\\

\subsection{Near horizon series expansions}
\label{app:nhseries}

As stated before, we will demand regularity of the solutions near the horizon. Thus, in the $u$ coordinate,
\be 
\left( \Phi , A \right) = \sum_{n=0} \left( \phi_H^{(n)}, A_H^{(n)} \right)(1-u)^n, \quad \left(f, A_0 \right) = \sum_{n=1} \left( f_H^{(n)}, A_0{}_H^{(n)} \right)\left(1-u\right)^n\ .
\ee
Again, combining this with the equations of motion, we obtain
\be \label{eq:nearhorseries}
\begin{split}
A_H^{(1)} & =  \frac{1}{f_H^{(1)}} \left[ \frac{1+ \phi_H^{(0)}{}^4}{ \left(\phi_H^{(0)}{}^2-1\right){}^2}-\frac{2}{3} A_{0H}^{(1)}{}^2 e^{-2 A_H^{(0)}}\right] -\frac{1}{2}  \\
A_H^{(2)} &= \frac{1}{f_H^{(1)} \left(\phi_H^{(0)}{}^2-1\right){}^2}\left[-\frac{4 A_{0H}^{(1)}{}^2 e^{-2 A_H^{(0)}} \phi_H^{(0)}{}^2}{3 f_H^{(1)}{}}+\frac{\phi_H^{(0)}{}^4}{2}+\frac{1}{2}\right] -\frac{A_{0H}^{(1)}{}^2e^{-2 A_H^{(0)}}}{3 f_H^{(1)}} \\
& -\frac{\phi_H^{(0)}{}^2}{2 f_H^{(1)}{}^2  \left(\phi_H^{(0)}{}^2-1\right){}^4}\left[\frac{3\phi_H^{(0)}{}^4}{2} + 3 \phi_H^{(0)}{}^2 + \frac{3}{2 }\right]  -\frac{1}{4} \\
A_0{}_H^{(2)} &= \frac{1}{6} A_{0H}^{(1)} \left(\frac{4 A_{0H}^{(1)}{}^2e^{-2 A_H^{(0)}}-6}{f_H^{(1)}}+3\right) 
\end{split}
\ee
and
\be \label{eq:nearhorseries2}
\begin{split}
f_H^{(2)} &= \frac{10}{3} A_{0H}^{(1)}{}^2 e^{-2 A_H^{(0)}}+\frac{f_H^{(1)}}{2}-\frac{2 \left(\phi_H^{(0)}{}^4+1\right)}{\left(\phi_H^{(0)}{}^2-1\right){}^2}  \\
\phi_H^{(1)} &= \frac{3 \left(\phi _H^{(0)}{}^2+1 \right)}{4 f_H^{(1)} \left(\phi _H^{(0)}{}^2-1\right)} \phi _H^{(0)} \\
\phi_H^{(2)} &= \frac{ \phi _H^{(0)} \left(\phi_H^{(0)}{}^2+1\right)}{64 f_H^{(1)}{}^2 \left(\phi _H^{(0)}{}^2-1\right){}^2}\Bigg\lbrace 3  \left[(8 f_H^{(1)}+9) \phi_H^{(0)}{}^2-8 f_H^{(1)}+3\right]-32 e^{2 A_H^{(0)}} A_{0H}^{(1)}{}^2 \left(\phi _H^{(0)}{}^2-1\right) \Bigg\rbrace 
\end{split}
\ee
plus higher-order terms.\\

\subsection{Numerical integration}
	\label{app:shooting}
We will solve the system of equations \Eq{eq:eoms} through the \emph{shooting technique} to determine the independent boundary and horizon constants. At given values $\left(\m_r,t_r\right)$, one starts with a trial set of independent boundary and horizon data, 
\be 
X= \left(A_0{}_H^{(1)},A_H^{(0)},A_0{}_{(0,2)},\alpha,\beta, \phi_H^{(0)}\right),
\ee
Note that $f_H^{(1)}$ can be fixed in terms of $t_r$ and $A_H^{(0)}$ alone and the constrain fixes the value of $A_H^{(1)}$. 

The algorithm is as follows: We compute the numerical solution and construct some object made out of the fields and their derivatives
\be 
V(u)= \left(f,A_0,\phi,A,A_0',\phi'\right),
\ee
note that it is not necessary to account for the derivatives of $f$ or $A_0$ since their equations of motion turn out to be first order. We perform the numerical integration from some near horizon value $u_{\text{hor}}$, using as boundary conditions the near-horizon series expansions from \Eq{eq:nearhorseries} and \Eq{eq:nearhorseries2}, down to some intermediate point $u_*$. Evaluating the fields and their derivatives at this point produces a vector $V(u_*)\vert_{\text{hor}\to \text{bulk}}$. Repeating the analogous procedure, this time employing the near-boundary series as boundary conditions, from some near-boundary value $u_{\text{boun}}$ down to the same intermediate point $u_*$ produces $V(u_*)\vert_{\text{boun}\to \text{bulk}}$.\footnote{Nevertheless, both for the near horizon and near boundary series expansions, in order to enhance the accuracy and shorten the overall integration time, we have truncated the series at a much larger order.} The mismatch vector $M$ is constructed by the difference
\be 
M(X) =  V(u_*)\vert_{\text{hor}\to \text{bulk}} -V(u_*)\vert_{\text{boun}\to \text{bulk}} \ .
\ee

The correct choice of $X$ must lead to $M = 0$. By thinking of $M(X)$ as a vector-valued function, the problem becomes a root finding in six dimensions. We apply the Newton-Raphson method. It works by a generalization of the familiar one-dimensional method of tracking tangent lines. For a guess $X$, compute the Jacobian $J$ of partial derivatives of the mismatch vector. The new vector $X$ shall be 
\be 
X = X^{\text{guess}} - J^{-1}M\ ,
\ee

The Jacobian is computed through finite differences, once the solutions in a neighborhood of the guess point (on each direction on the constants space) are known. In particular, as step in the Jacobian we will take $10^{-10}$.
On each numerical integration, $u_{\text{hor}}=1-\epsilon_0$, $u_{\text{boun}}=\epsilon_0$, $\epsilon_0$ being some sensitive cut-off; we use $10^{-8}$ and $u_*=1/2$. As for the initial data $X^{\text{guess}}$, a sensitive choice for mild reduced chemical potential and temperature is the solution inherited from the scalar field in probe approximation, wherein the geometry reduced to an AdS-RN\cite{Hoyos:2016cob},
\be 
X^{\text{guess}} = X_{\text{AdS-RN}} = \left(\mu,0,0,\alpha^P,\beta^{P},\phi_H^P \right) \ , 
\ee
where $\left(\alpha^P,\beta^P\right)$ are obtained from integration of the scalar equation in this approximation, once $\phi_H^P$ is set. If the norm of the mismatch $|| M ||$ lies above some threshold fixed a priori, the iteration starts once again, but taking $X$ as the new starting point and stops if otherwise. In our computations, we will fix the threshold to be $10^{-9}$. Our attempt to connect the model to neutron star physics implies that we will focus in regimes at which $X_{\text{AdS-RN}}$ works not very well, but luckily, thanks to the smoothness of the solutions, if for some choice $X_{\left(\m_0, t_0\right)}, || M || < 10^{-9}$, then we can take this vector as initial guess on the next computation, i.e., $X_{\left(\m_0, t_0\right)}\to X^{\rm guess}_{\left(\m_0+\delta \mu, t_0\right)}$.

\section{Calculation of thermodynamic quantities}\label{app:thermo}

\subsection{On-shell action}\label{app:onshell}
%%%%%%%%%%%%%%%%%%%%%%%%%%%%%

For the holographic models we consider, one can write Einstein's equations in the form
\begin{equation}
R_{MN} = T_{MN}^{(A)}+T_{MN}^\phi+\frac{1}{2} g_{MN}\left (\frac{L^2}{3}F^2+\mathcal{K}_\Phi |D\phi|^2+\frac{5}{3}\mathcal{V}_\Phi\right) \ .
\end{equation}
From the trace of these equations, we find that the Ricci scalar reads
\begin{equation}
R=\frac{L^2}{3}F^2+\mathcal{K}_\Phi|D\phi|^2+\frac{5}{3}\mathcal{V}_\Phi \ ,
\end{equation}
implying that the on-shell action \eqref{eq:langdens} evaluates to
\begin{equation}
S_{\rm on-shell}=\frac{1}{16\pi G_5}\int d^5 x\sqrt{-g}\left[\frac{2}{3}\mathcal{V}_\Phi-\frac{2}{3}L^2 F^2 \right] \ .
\end{equation}
Let us now use the fact that for our solutions
\begin{equation}
\Gamma^\alpha_{\mu\nu}=\Gamma^r_{r\nu}=\Gamma^\alpha_{rr}=0 \ , \ \Gamma^r_{\mu\nu}=-\frac{1}{\sqrt{g_{rr}}} K_{\mu\nu}\ , \ \Gamma^\alpha_{\mu r}=\sqrt{g_{rr}} K^\alpha_{\ \mu}\ , \ \Gamma^r_{rr}=\frac{1}{2}g^{rr}\partial_r g_{rr} \ ,
\end{equation}
where 
\begin{equation}
K_{\mu\nu}=\frac{1}{2\sqrt{g_{rr}}}\partial_r g_{\mu\nu} \
\end{equation}
is the extrinsic curvature and $K^\alpha_{\ \mu}=g^{\alpha\beta}K_{\beta\mu}$, $K=g^{\mu\nu}K_{\mu\nu}$. Using also  the simple result
\begin{equation}
\frac{\partial_r \sqrt{-g}}{\sqrt{-g}}=\Gamma^r_{rr}+\sqrt{g_{rr}}K\ ,
\end{equation}
we can write
\begin{equation}
g^{\mu\nu}R_{\mu\nu}=-\frac{1}{\sqrt{-g}}\partial_r\left( \frac{\sqrt{-g}}{\sqrt{g_{rr}}}K\right)=-\frac{1}{\sqrt{-g}}\partial_r\left( \sqrt{-\gamma} K\right) \ .
\end{equation}
Here, we defined $\gamma_{\mu\nu}=g_{\mu\nu}$ as the boundary metric and used $\sqrt{-g}=\sqrt{g_{rr}}\sqrt{-\gamma}$.

On the other hand, from Einstein's equations we obtain
\begin{equation}
g^{\mu\nu} R_{\mu\nu}=-\frac{2L^2}{3}F_{0 r} F^{0 r}+\frac{4}{3}\mathcal{V}_\Phi +q^2 g^{00} \mathcal{K}_\Phi A_0^2\phi^2 \ ,
\end{equation}
where we only focused on the nonzero components of the solutions. Solving now for $\mathcal{V}_\Phi$ and introducing the result in the on-shell action, one gets
\begin{equation}
S_{\rm on-shell}=\frac{1}{16\pi G_5}\int d^5 x\sqrt{-g}\left[-\frac{1}{2\sqrt{-g}}\partial_r\left( \sqrt{-\gamma}K\right)-L^2 F_{r0}F^{r0}-\frac{q^2}{2} g^{00} \mathcal{K}_\phi A_0^2 \phi^2 \right]\ .
\end{equation}
Finally, we use the equation of motion for the gauge field,
\begin{equation}
4L^2\partial_r\left( \sqrt{-g} F^{r0}\right)=2 q^2 \sqrt{-g} g^{00} \mathcal{K}_\phi A_0\phi^2 \ .
\end{equation}
We can then replace the $q^2$ term in the action by a derivative term and write the action as a total derivative:
\bea
S_{\rm on-shell} & = & \frac{1}{16\pi G_5}\int d^5 x\left[-\frac{1}{2}\partial_r\left(\sqrt{-\gamma}K\right)-L^2 \sqrt{-g} \partial_r A_0 F^{r0}-L^2A_0\partial_r\left( \sqrt{-g} F^{r0}\right)  \right]\nonumber\\
 & = & \frac{1}{16\pi G_5}\int d^5 x \,\partial_r \left[-\frac{1}{2}\sqrt{-\gamma}K-L^2 \sqrt{-g}  A_0 F^{r0} \right] \nonumber\\
 & = & \frac{1}{16\pi G_5} \int d^4 x \left[ \frac{1}{2} \sqrt{-\gamma}K+L^2 \sqrt{-g}  A_0 F^{r0}\right]_{r=r_H}^{r=r_\Lambda} \ . \label{eq:onshellact}
\eea

\subsection{Holographic renormalization}
\label{app:holoRG}

In order to be able to read off the speed of sound, we need the energy density $\varepsilon$ and pressure $p$, which can be read from the diagonal components of the expectation value of the stress energy tensor, $\vev{T_{\m\n}}$. We can decompose the line element \Eq{eq:holomet} into its transverse and longitudinal components,
\be 
dS^2 = N^2 dr^2 + \gamma_{\m\n}dx^\m dx^\n\,, \qquad N^2= \frac{L^2}{r^2 f}\,.
\ee
We will now determine, which counterterms we need to consider in order to obtain finite one point correlation functions. Together with the cosmological constant term
\be 
\mathcal{I}_\Lambda = -\frac{1}{8 \pi G_5}\int d^4 x\sqrt{-\gamma}\Lambda \, \,, %\quad d=4\,,
\ee
which will cancel out the volume divergence, we need to include also the Gibbons-Hawking term, 
\be 
\mathcal{I}_{\rm GH} = \frac{1}{8 \pi G_5}\int d^4 x\sqrt{-\gamma}K \, \,. %\quad d=4\,,
\ee
The details of the holographic renormalization of bottom-up models can be found in \cite{Hoyos:2016cob}. In the following we focus on the top-down model, that present some small differences due to the more complicated form of the kinetic term and the potential for the scalar field.

From the near boundary behavior of the metric field,
\be 
\begin{split}
	\gamma_{00} &= -\frac{r^2}{L^2} + \frac{4}{3}L^2 \phi_{(0,1)}^2 + \frac{L^6}{9r^2}\Bigg\lbrace  \left[2 \phi_{(0,1)} \left(9 \mu ^2 \phi_{(0,1)}+\phi_{(0,1)}^3+9 \phi_{(0,3)}\right)-9f_{0,4}\right]\\
	& + 12 \phi_{(0,1)}^2 \left(15 \mu ^2+2\phi_{(0,1)}^2\right) \log \left(\frac{r}{L}\right)\Bigg\rbrace  \\
	\gamma_{ii} &= \frac{r^2}{L^2}  -\frac{4}{3}L^2\phi_{(0,1)}^2 -\frac{2 L^6}{9 r^2} \Bigg\lbrace  \phi_{(0,1)} \left(9 \mu ^2 \phi_{(0,1)}+\phi_{(0,1)}^3+9\phi_{(0,3)}\right) \\
	& + 6\phi_{(0,1)}^2 \left(3 \mu ^2+2\phi_{(0,1)}^2\right) \log \left(\frac{r}{L}\right)\Bigg\rbrace \ ,
\end{split}
\ee
we note that it is necessary to add the following counterterm that will cancel out divergences due to the backreaction of the scalar field,
\be 
\mathcal{I}_{\text{c}} = -\frac{1}{8\pi G_5}\int d^4 x\, \sqrt{-\gamma} \Bigg\lbrace \frac{32}{4}L\vert D \Phi\vert^2 \log\left(\frac{r}{L}\right) -\left[8+ \frac{32}{3}\Phi^2 \log\left(\frac{r}{L}\right)\right]\frac{\Phi^2}{L} \Bigg\rbrace \ .
\ee 
Another counterterm may also be added,
\be
\mathcal{I}_{\text{f}} = -\frac{L}{8\pi G_5} \int d^4 x \, \sqrt{-\gamma} \left[ \mathcal{W}_1 \vert D_\a \Phi\vert^2 + \frac{\mathcal{W}_2}{L^2} \Phi^4 \right]\ ,
\ee
which will introduce non-trivial finite contributions to our QFT.

% After varying the action with respect to the boundary metric, The renormalized stress-tensor components read CHECK
% \bea
% \vev{T^{\m\n}} &=& -\frac{1}{8\pi G_5}\lim_{r\to \infty}\frac{r^2}{L^2} e^{2A}\sqrt{-\gamma} \times \Bigg\lbrace \Pi^{\m\n}_{BY} +\Pi^{\m\n}_{GH+\Phi^2} + \nonumber\\
% & &  \Pi^{\m\n}_{\mathcal{D}\phi}\left[\log \left(\frac{r}{L}\right) + \mathcal{W}_1 \right]  + \Pi^{\m\n}_{\Phi^4} \left[\log \left(\frac{r}{L}\right) + \mathcal{W}_2\right]  \Bigg\rbrace,\\
% \vev{\mathcal{O}} &=& \frac{1}{8\pi G_5}\lim_{r\to \infty} \times \Bigg( \sqrt{-g}g^{rr}\partial_r\Phi +\\
% & &  \frac{1}{L}\sqrt{-\gamma}\Bigg\lbrace \Phi -\left[ \log \left(\frac{r}{L}\right) + \mathcal{W}_1\right] L^2\mathcal{D}^2 \Phi -\left[\frac{2}{3} \log \left(\frac{r}{L}\right) + 2\mathcal{W}_2 \right] \Phi^3  \Bigg\rbrace \Bigg),
% %\vev{J^\mu} &=& -\frac{R}{16\pi G_5 g^2} \lim_{r\to \infty} \sqrt{-g} g^{rr} g^{\mu \alpha}F_{r\alpha},
% \eea
% \be
% \begin{split}
% 	\Pi^{\m\n}_{\mathcal{D}\Phi} &= -\frac{L}{2}\Bigg\lbrace \left[\left(\mathcal{D}^\m \Phi \right)^*\mathcal{D}^\n\Phi +\left(\mathcal{D}^\n \Phi \right)^*\mathcal{D}^\m\Phi\right] -\gamma^{\m\n}(\mathcal{D}_\a\Phi)^*\mathcal{D}^\a\phi \Bigg\rbrace,  \\
% 	\Pi^{\m\n}_{\Phi^4} &=  \frac{1}{3L}\gamma^{\m\n}\Phi^4 .
% \end{split}
% \ee

After varying the action with respect to the boundary metric, and inserting the near boundary series expansions \Eq{eq:nbseries}, we get the boundary vev's
\be \label{eq:emtvevs}
\begin{split} 
\vev{T^{00}} = \varepsilon &= -\frac{L^3}{16\pi G_5}\left[ 3 f_{(0,4)}+8 \phi_{(0,1)}\phi_{(0,3)} +4 \mu^2 \phi_{(0,1)}^2 (\mathcal{W}_1+3)+\phi_{(0,1)}^4\left(\mathcal{W}_2+\frac{16}{3}\right) \right] \\
\vev{T^{ii}} = p &= -\frac{L^3}{16\pi G_5}\left[ f_{0,4} - 8\phi_{(0,1)}\phi_{(0,3)} + 4 \mu ^2 \phi_{(0,1)}^2 (\mathcal{W}_1 +1) -\phi_{(0,1)}^4 \left(\mathcal{W}_2 +\frac{16}{3}\right) \right] \\
\vev{\mathcal{O}} = v &= -\frac{2L^3}{\pi G_5}\left[\phi_{(0,3)} -\frac{1}{4} \mu ^2 \phi_{(0,1)} (\mathcal{W}_1 +4)+\phi_{(0,1)}^3
   \left(\frac{\mathcal{W}_2}{8}-\frac{2}{3}\right)\right]  \\
\vev{j^0} = n &= -\frac{L^3}{2\pi G_5}\left[A_0{}_{(0,2)}+\mu  \phi_{(0,1)}^2 (\mathcal{W}_1+4)\right],
\end{split} 
\ee
which satisfy
\be 
\vev{T^{\mu\nu}}\eta_{\mu\nu} = - \vev{\mathcal{O}}\phi_{(0,1)} + \mathcal{A} \ ,
\ee
with the anomaly
\be 
\mathcal{A} = \frac{L^3 }{\pi  G_5}  \phi_{(0,1)}^2\left(\frac{\mu ^2}{2} +\frac{2}{ 3} \phi_{(0,1)}^2\right)\ . 
\ee

Combining expressions \Eq{eq:emtvevs} and \Eq{eq:tement}, one can straightforwardly verify that the thermodynamic relation
\be 
\varepsilon + p = n \mu + T S\ 
\ee
holds. Moreover, the renormalized action at the boundary is equal to the free energy in the macrocanonical ensemble,
\be 
\begin{split}
S_{\rm ren} &= S_{\rm on-shell} +\mathcal{I}_{\Lambda} + \mathcal{I}_{\rm GH} + \mathcal{I}_{\rm c} + \mathcal{I}_{\rm f}   \\
&= \Omega =  \frac{L^3}{16\pi G_5}\left[ f_{0,4} - 8\phi_{(0,1)}\phi_{(0,3)} + 4 \mu ^2 \phi_{(0,1)}^2 (\mathcal{W}_1 +1) -\phi_{(0,1)}^4 \left(\mathcal{W}_2 +\frac{16}{3}\right) \right] \\
&= -p \ ,
\end{split}
\ee
where we have made use of \Eq{eq:rel1} when expressing $S_{\rm on-shell}$ at the horizon $r_H$ in terms of the boundary coefficients.

We can now examine the equations of motion in order to see if some sort of relation between the near boundary/horizon coefficients can be set. If we define
\begin{equation}\label{eq:bg}
\beta(r)=e^{4A} r^5 f'-4L^4  e^{2A} r^3 A_0' A_0\ ,
\end{equation}
we notice that due to equations \Eq{eq:eoms}, this quantity is independent of the radial coordinate. It is convenient to evaluate it at the horizon, $r\to r_H$, giving
\bea
\beta(r_H)=e^{4A(r_H)} r_H^5 f'(r_H) & \equiv & \beta_H  \ . \label{eq:bgH}
\eea
Note also that the temperature and entropy density are given by
\begin{equation} \label{eq:tement}
T=\frac{r_H^2 f'(r_H) }{4\pi L^2}e^{A(r_H)} \ , \ s=\frac{1}{4G_5} \frac{r_H^3}{L^3}e^{3 A(r_H)} \ ,
\end{equation}
so that
\begin{equation}\label{eq:betaH}
\beta_H=16\pi G_5  L^5 T s \ .
\end{equation}

The above steps enable us to find the relation 
\be \label{eq:rel1}
\widehat{f}_{(0)} = \alpha \left(2 \mu_r a_1 +4 \mu_r^2 \alpha^3-\pi  e^{3 A_H^{(0)}} t_r\right).
\ee 
Moreover, another relation can be obtained from the constraint equation in the bulk,
\be \label{eq:rel2}
A_H^{(1)} =  \frac{1}{f_H^{(1)}} \left[ \frac{1+ \phi_H^{(0)}{}^4}{ \left(\phi_H^{(0)}{}^2-1\right){}^2}-\frac{2}{3} A_{0H}^{(1)}{}^2 e^{-2 A_H^{(0)}}\right] -\frac{1}{2}\ .
\ee 
Both relations \Eq{eq:rel1} and \Eq{eq:rel2} can be employed to enhance the numeric integration of the set of equations \Eq{eq:eoms}.

\section{Fluctuations}
\label{app:numerics}

\subsection{Equations for gauge invariant combinations}

We will use radial gauge $\delta g_{\mu r}=\delta a_r=0$. At zero spatial momentum fluctuations split in decoupled sectors according to their representation under the group of spatial rotations. There are three sectors:
\begin{itemize}
\item Tensor: $h_{ij}-\frac{1}{3}\delta_{ij}\delta^{kl}h_{kl}$ \ .  
\item Vector: $a_i$, $h_{0i}$\ .
\item Scalar: $\varphi$, $\varphi^\dagger$, $h_{00}$, $a_0$, $h=\delta^{ij}h_{ij}/3$\ . 
\end{itemize}
In principle we expect instabilities to be related to changes in the scalar, thus we will restrict the analysis to the scalar sector. We see that there are five components of the fields in the scalar sector. The equations of motion (Einstein, Maxwell, and the equation of motion for the scalar) include a second order (dynamical) equation for each mode plus three first order (constraints) equations. This adds up to eight coupled equations for the five modes. However, the actual number of independent dynamical modes is just two and the system can be reduced to two coupled differential equations (of second order). We will do this in the following.

In the radial gauge there are residual diffeomorphisms $\xi^M(x)$ and gauge transformations $\lambda(x)$. The linear variations of the fields are
\be
\begin{split}
\delta \Phi = & \xi^M\partial_M\Phi+i q \Phi \lambda \\
\delta \Phi^\dagger = & \xi^M\partial_M\Phi^\dagger-i q \Phi^\dagger \lambda \\
\delta A_M = & \xi^N\partial_N A_M+\partial_M\xi^N A_N+\partial_M \lambda\ .
\end{split}
\ee
For homogeneous fluctuations we can expand in plane waves $\xi^M=e^{-i\omega t}\eta^M(r)$, $\lambda=e^{-i\omega t}\chi(r)$, in such a way that the allowed transformations are
\be
\eta^r=c_0 r\sqrt{f} , \ {\eta^0}'=-i\omega c_0 \frac{L^4e^{-2A}}{r^3 f^{3/2}},\ \ \chi' =i\omega c_0  \frac{L^4e^{-2A}}{r^3 f^{3/2}}A_0,\ \ \eta^i=c_i\ ,
\ee
where $c_0$, $c_i$ are arbitrary functions of the frequency. We can construct a basis of two independent combinations of the scalar components that are invariant under these gauge transformations $z_1,z_2$; these are the expressions given in \eqref{eq:zs}. The equations of motion can be found in a straightforward way by taking radial derivatives of $z_i$ and using the equations of motion of the scalar modes. They take the generic form \eqref{eq:eomsfluc}. The result with $q\neq0$ is quite cumbersome, so we will give here expressions for the bottom-up models with $q=0$ and canonical kinetic term, but generic potential. The off-diagonal coefficients vanish $\cA_{12}=\cA_{21}=\cB_{12}=\cB_{21}=0$ and the diagonal ones take the values:
 \be
\begin{split}
\cA_{11}=\cA_{22}=&-4 A'-\frac{f'}{f}-\frac{5}{r}\\
\cB_{11}=& -\frac{e^{-2 A} L^4 \omega ^2}{f^2 r^4}+\frac{4 \Phi _0 \Phi _0'\partial  \cV_{\Phi }}{3 f r^2 A'+3 f r}-\frac{2 r f' \left(\Phi _0'\right){}^2}{3 f \left(r
   A'+1\right)}-\frac{8}{3} \left(\Phi _0'\right){}^2\\
	&+\frac{\partial\cV_{\Phi }+2 \Phi _0^2\partial^2  \cV_{\Phi }}{f r^2}+\frac{2 r^2 \left(\Phi _0'\right){}^4}{9 \left(r A'+1\right)^2}\\
	\cB_{22}=& \frac{\partial\cV_{\Phi }}{f r^2}-\frac{e^{-2 A} L^4 \omega ^2}{f^2 r^4}\ .
\end{split}
\ee

\subsection{Solutions}

The method that we will follow here to find a solution for the quasi-normal modes is valid for any number of coupled or decoupled linear differential equations. Expanding the system \Eq{eq:eomsfluc} around $u\to 1$,
\be 
0 = z_j'' - \frac{z_j'}{1-u}  + \frac{e^{-2A_H^{(0)}}\omega^2}{4 f_H^{(1)} (1-u)^2} z_j \ , 
\ee
we infer that the leading order behavior at the horizon is given by
\be 
z_j\vert_{u\to 1} \sim z_j^{(\text{out})}(1-u)^{i\omega c_I} + z_j^{(\text{ing})}(1-u)^{-i\omega c_I}
\ee
with $c_I=e^{A_H^{(0)}} /2 f_H^{(1)}$, and we have labeled the outgoing and infalling pieces as $z_j^{(\text{out})}$ and $z_j^{(\text{ing})}$, respectively. Imposing causality means that we pick the ingoing solution. From here, we can construct a solution valid throughout the whole bulk,
\be 
z_j \sim (1-u)^{-i\omega c_I}  z_j{}_{(\text{reg})}\ ,
\ee
with 
\be \label{eq:zreg}
z_j{}_{(\text{reg})} = \sum_{m=0} z_j^{(m)}(1-u)^m\ ,
\ee
regular at the horizon. At leading order and taking $\omega=0$,
\be 
\begin{split}
z_1{}_{(\rm reg)} &= z_1^{(1)}(1-u) + \cdots \\
z_2{}_{(\rm reg)} &= z_2^{(0)} - \Bigg\lbrace \frac{\left(\phi_H^{(0)}{}^2-1\right) \left(\phi_H^{(0)}{}^2+1\right) }{2 \left(\phi_H^{(0)}{}^4+1\right)f_H{}^{\text{(0)}}}z_1{}^{\text{(1)}} e^{-2A_H{}^{\text{(0)}}} A_0{}_H{}^{\text{(1)}}\\
& +\frac{3 \left(\phi_H^{(0)}{}^8+8 \phi_H^{(0)}{}^4-1\right) z_2{}^{\text{(0)}}}{4 \left(\phi_H^{(0)}{}^2-1\right){}^2 \left(\phi_H^{(0)}{}^4+1\right) f_H{}^{\text{(0)}}} \Bigg\rbrace (1-u) +\cdots \ .
\end{split}
\ee

A normalizable solution at $u\to 0$ can be obtained by means of the determinant method. First, we choose a set of linearly independent boundary conditions at the horizon, that is,
\be 
\big\lbrace z_1{}_{(reg)}, z_2{}_{(\text{reg})} \big\rbrace = \big\lbrace \left(1,0\right), \left(0,1\right)\big\rbrace\ ,
\ee
and for each of these boundary conditions, we solve numerically the system \Eq{eq:eomsfluc} by means of a single shooting from the horizon, where we impose
\be 
z_j( 1-\epsilon_0)=z_j{}_{(\rm reg)}( 1-\epsilon_0), \quad z_j'( 1-\epsilon_0) =z_j{}_{(\rm reg)}{}'( 1-\epsilon_0)\ ,
\ee
to the boundary, taking as cutoff the same as in the background computation ($\epsilon_0 = 10^{-8}$), although there is a high robustness against this choice. Furthermore, since we now deal with a linear differential equation system, there is no need to demand the same accuracy as for the background computation, so we set $m=2$ in Eq.\Eq{eq:zreg}. Near the boundary, the solutions have the following expansion to leading order, 
\be 
z_{1,2} \sim \sqrt{u}\left(z_{1,2}^{(nn)} + u\, z_{1,2}^{(n)} \right)\ ,
\ee 
where we identify the non-normalizable $(nn)$ as the leading term while the normalizable $(n)$ as the sub-leading one. Normalizable solutions will have $z_{1}^{(nn)}=z_{2}^{(nn)}= 0$. 
The numerical solutions can be arranged as elements of a matrix $M$,
\be 
M= \frac{1}{\sqrt{u}}\left(
\begin{array}{c c}
z_1^{(I)}{}_{(\text{reg})} & z_2^{(I)}{}_{(\text{reg})}\\
z_1^{(II)}{}_{(\text{reg})}  & z_2^{(II)}{}_{(\text{reg})}\\
\end{array}
\right)\ ,
\ee
which, if evaluated at the $AdS$ boundary gives zero determinant, then, a normalizable solution exists. This will happen at a certain frequency $\omega \in \mathbb{C}$, for fixed chemical potential and temperature. If we were about to determine such frequency, the problem amounts to find the root of a certain equation, $\text{det}(M(\omega))=0$, which can be searched using Newton's method. Nevertheless, this might not even be necessary, since we can dial the chemical potential and compute the determinant at zero frequency.

\bibliographystyle{JHEP}

\bibliography{biblio3}

\end{document}